\documentclass[12pt]{article}%
\usepackage{ amsmath, graphics, rotating}%

\begin{document}

\begin{center}{\large\bf Introduction to a Quantum Theory over a Galois Field}\end{center}
\vskip 1em \begin{center} {\large Felix M. Lev} \end{center}
\vskip 1em \begin{center} {\it Artwork Conversion Software Inc.,
1201 Morningside Drive, Manhattan Beach, CA 90266, USA
(Email:  felixlev314@gmail.com)} \end{center}
\vskip 1em

\begin{flushleft}{\it Abstract:}\end{flushleft} 
We consider a quantum theory based on a Galois field.
In this approach infinities cannot exist, the cosmological
constant problem does not arise, and one irreducible
representation (IR) of the symmetry algebra splits into
independent IRs describing a particle an its antiparticle only
in the approximation when de Sitter energies are much less than
the characteristic of the field. As a consequence, the very
notions of particles and antiparticles are only approximate and
such additive quantum numbers as the electric, baryon and
lepton charges are conserved only in this approximation. There
can be no neutral elementary particles and the spin-statistics
theorem can be treated simply as a requirement that standard
quantum theory should be based on complex numbers. 

\begin{flushleft}{Key words: quantum theory; Galois fields; elementary particles}\end{flushleft} 
\begin{flushleft}{PACS: 02.10.De, 03.65.Ta, 11.30.Fs, 11.30.Ly}\end{flushleft}

\section{Motivation}
\label{S1}

The most striking feature of the modern quantum theory is
probably the following. On one hand, this theory describes many
experimental data with an unprecedented accuracy. On the other
hand, the mathematical substantiation of the theory is rather
poor. As a consequence, the issue of infinities is probably the
most challenging problem in standard formulation of quantum
theory. As noted by Weinberg \cite{Wein}, {\it
``Disappointingly this problem appeared with even greater
severity in the early days of quantum theory, and although
greatly ameliorated by subsequent improvements in the theory,
it remains with us to the present day''}. While in QED and
other renormalizable theories infinities can be somehow
circumvented, in quantum gravity this is not possible even in
lowest orders of perturbation theory. A recent Weinberg's paper
\cite{Wein-inf} is entitled ``Living with Infinities''.

Mathematical problems of quantum theory are discussed in a wide
literature. For example, in the well known textbook
\cite{Bogolubov} it is explained in details that interacting
quantized fields can be treated only as operatorial
distributions and hence their product at the same point is not
well defined. One of ideas of the string theory is that if a
point (a zero-dimensional object) is replaced by a string (a
one-dimensional object) then there is a hope that infinities
will be less singular.

There exists a wide literature aiming to solve the difficulties
of the theory by replacing the field of complex numbers by
quaternions, p-adic numbers or other constructions. For
example, a detailed description of a quaternionic theory is
given in a book \cite{Adler} and a modern state-of-the-art of
the p-adic theory can be found, for example, in Reference
\cite{Dragovich}. At present it is not clear how to overcome
all the difficulties but at least from the point of view of the
problem of infinities a natural approach is to consider a
quantum theory over Galois fields (GFQT). Since any Galois
field is finite, the problem of infinities in GFQT does not
exist in principle and all operators are well defined. The idea
of using finite fields in quantum theory has been discussed by
several authors (see e.g., References 
\cite{Galois,Shapiro,Nambu,Vourdas1,Doughty,Vourdas2,Volovich}). 
As stated in Reference \cite{Volovich}, a fundamental theory can
be based either on p-adic numbers or finite fields. In that
case, a correspondence with standard theory will take place if
the number $p$ in the p-adic theory or as a characteristic of a
finite field is rather large.

The authors of Reference \cite{Volovich} and many other papers
argue that fundamental quantum theory cannot be based on
mathematics using standard geometrical objects (such as
strings, branes, \emph{etc.}) at Planck distances. We believe
it is rather obvious that the notions of continuity,
differentiability, smooth manifolds \emph{etc.} are based on
our macroscopic experience. For example, the water in the ocean
can be described by equations of hydrodynamics but we know that
this is only an approximation since matter is discrete.
Therefore continuous geometry probably does not describe
physics even at distances much greater than the Planck length
(also see the discussion below).

In our opinion an approach based on finite fields is very
attractive  for solving problems in quantum theory as well as
for philosophical and aesthetical reasons. Below we describe
some arguments in favor of this opinion.

The key ingredient of standard mathematics is the notions of
infinitely small and infinitely large numbers. The notion of
infinitely small numbers is based on our everyday experience
that any macroscopic object can be divided by two, three and
even a million parts. But is it possible to divide by two or
three the electron or neutrino? It is obvious that if
elementary particles exist, then division has only a limited
meaning. Indeed, consider, for example, the gram-molecule of
water having the mass 18 grams. It contains the Avogadro number
of molecules $6\cdot 10^{23}$. We can divide this gram-molecule
by ten, million, \emph{etc.}, but when we begin to divide by
numbers greater than the Avogadro one, the division operation
loses its sense. The conclusion is that {\it if we accept the
existence of elementary particles, we should acknowledge that
our experience based on standard mathematics is not universal}.

The notion of infinitely large numbers is based on the belief
that {\it in principle} we can operate with any large numbers.
In standard mathematics this belief is formalized in terms of
axioms (accepted without proof) about infinite sets (e.g.,
Zorn's lemma or Zermelo's axiom of choice). At the same time,
in the spirit of quantum theory, there should be no statements
accepted without proof since only those statements have
physical significance, which can be experimentally verified, at
least in principle.

For example, we cannot verify that $a+b=b+a$ for any numbers
$a$ and $b$. Suppose we wish to verify that 100+200=200+100. In
the spirit of quantum theory, it is insufficient to say that
100+200=300 and 200+100=300. To check these relationships, we
should describe an experiment where they can be verified. In
particular, we should specify whether we have enough resources
to represent the numbers 100, 200 and 300. We believe the
following observation is very important: although standard
mathematics is a part of our everyday life, people typically do
not realize that {\it standard mathematics is implicitly based
on the assumption that one can have any desirable amount of
resources}.

A well known example in standard mathematics is that the
interval $(0,1)$ has the same cardinality as
$(-\infty,\infty)$. Another example is that the function $tgx$
gives a one-to-one relation between the intervals $(-\pi /2,\pi
/2)$ and $(-\infty,\infty)$. Therefore one can say that a part
has the same number of elements as a whole. One might think
that this contradicts common sense but in standard mathematics
the above facts are not treated as contradicting.
Self-consistency of standard mathematics has been discussed by
numerous authors. For example, the famous Goedel's
incompleteness theorems are interpreted as showing that
Hilbert's program to find a complete and consistent set of
axioms for all of mathematics is impossible.

Suppose now that our Universe is finite and contains only a
finite number of elementary particles. This implies that the
amount of resources cannot be infinite and the rules of
arithmetic such as $a+b=b+a$ for any numbers $a$ and $b$,
cannot be verified in principle. In this case it is natural to
assume that there exists a number $p$ such that all
calculations can be performed only modulo $p$. Note that for
any system with a finite amount of resources, the only way of
performing self-consistent calculations is to perform them
modulo some number. One might consider a quantum theory over a
Galois field with the characteristic $p$. Since any Galois
field is finite, the fact that arithmetic in this field is
correct can be verified, at least in principle, by using a
finite amount of resources. Note that the proofs of the Goedel
incompleteness theorems are based on the fact that standard
arithmetic is infinite but in our case no inconsistencies
arise.

The example with division might be an indication that, in the
spirit of Reference \cite{Planat}, the ultimate quantum theory
will be based even not on a Galois field but on a finite ring
(this observation was pointed out to me by Metod Saniga).
However, in the present paper we will consider a case of Galois
fields.

If one accepts the idea to replace complex numbers by a Galois
field, the problem arises what formulation of standard quantum
theory is most convenient for that purpose. A well known
historical fact is that originally quantum theory has been
proposed in two formalisms which seemed essentially different:
the Schroedinger wave formalism and the Heisenberg operator
(matrix) formalism. It has been shown later by Born, von
Neumann and others that both formalisms are equivalent and, in
addition, the path integral formalism has been developed.

In the spirit of the wave or path integral approach one might
try to replace classical spacetime by a finite lattice which
may even not be  a field. In that case the problem arises what
the natural ``quantum of spacetime'' is and some of physical
quantities should necessarily have the field structure. A
detailed discussion can be found in Reference 
\cite{Galois,Shapiro,Nambu,Vourdas1,Doughty,Vourdas2} and
references therein. In contrast to these approaches, we propose
to generalize the standard operator formulation, where quantum
systems are described by elements of a projective complex
Hilbert spaces and physical quantities are represented by
self-adjoint operators in such spaces.

From the point of view of quantum theory, any physical quantity 
can be discussed only in conjunction
with the operator defining this quantity. However, in textbooks
on quantum mechanics it is usually not indicated explicitly
that the quantity $t$ is a parameter, which has the meaning of
time only in the classical limit since there is no operator
corresponding to this quantity. The problem of how time should
be defined on quantum level is very difficult and is discussed
in a vast literature (see e.g., References \cite{Rosen,Rosen2,Rickles2}
and references therein). Since the 1930's it has been well
known \cite{NW} that, when quantum mechanics is combined with
relativity, there is no operator satisfying all the properties
of the spatial position operator. In other words, the
coordinate cannot be exactly measured even in situations when
exact measurement is allowed by the non-relativistic
uncertainty principle. In the introductory section of the
well-known textbook \cite{BLP} simple arguments are given that
for a particle with mass $m$, the coordinate cannot be measured
with the accuracy better than the Compton wave length
${\hbar}/mc$. Hence, the exact measurement is possible only
either in the non-relativistic limit (when $c\to\infty$) or
classical limit (when ${\hbar}\to 0)$. From the point of view
of quantum theory, one can discuss if the {\it coordinates of
particles} can be measured with a sufficient accuracy, while
the notion of empty spacetime background fully contradicts
basic principles of this theory. Indeed, the coordinates of
points, which exist only in our imagination are not measurable
and this problem has been discussed in a wide literature (see
e.g., References \cite{Rosen,Rosen2,Rickles2,Rickles1}). In particular,
the quantity $x$ in the Lagrangian density $L(x)$ is not
measurable. Note that Lagrangian is only an auxiliary tool for
constructing Hilbert spaces and operators and this is all we
need to have the maximum possible information in quantum
theory. After this construction has been done, one can safely
forget about Lagrangian and concentrate his or her efforts on
calculating different observables. As stated in Reference
\cite{BLP}, local quantum fields and Lagrangians are
rudimentary notion, which will disappear in the ultimate
quantum theory. Analogous ideas were the basis of the
Heisenberg S-matrix program.

In view of the above discussion, we define GFQT as a theory where
\begin{itemize}
\item {\it Quantum states are represented by elements of a linear projective space over a Galois field
and physical quantities are represented by linear operators in that space.}
\end{itemize}

As noted in Reference \cite{Dragovich} and references therein,
in the p-adic theory a problem arises what number fields are
preferable and there should be quantum fluctuations not only of
metrics and geometry but also of the number field. Volovich
\cite{Volovich} proposed the following number field invariance
principle: fundamental physical laws should be invariant under
the change of the number field. Analogous questions can be
posed in GFQT.

It is well known (see, e.g., standard textbooks \cite{VDW,Ireland,Davenport})
that any Galois field can contain only $p^n$ elements where $p$
is prime and $n$ is natural. Moreover, the numbers $p$ and $n$
define the Galois field up to isomorphism. It is natural to
require that there should exist a correspondence between any
new theory and the old one, \emph{i.e.}, at some conditions both
theories should give close predictions. In particular, there
should exist a large number of quantum states for which the
probabilistic interpretation is valid. Then, as shown in our
papers \cite{lev2,lev22,hep}, in agreement with References
\cite{Galois,Shapiro,Nambu,Vourdas1,Doughty,Vourdas2,Volovich}, 
the number $p$ should be very large.
Hence, we have to understand whether there exist deep reasons
for choosing a particular value of $p$ or it is simply an
accident that our Universe has been created with this value.
Since we don't know the answer, we accept a simplest version of
GFQT, where there exists only one Galois field with the
characteristic $p$, which is a universal constant for our
Universe. Then the problem arises what the value of $n$ is.
Since there should exist a correspondence between GFQT and the
complex version of standard quantum theory, a natural idea is
to accept that the principal number field in GFQT is the Galois
field analog of complex numbers which is constructed below.

Let $F_p=Z/pZ$ be a residue field modulo $p$ and $F_{p^2}$ be a
set of $p^2$ elements $a+bi$ where $a,b\in F_p$ and $i$ is a
formal element such that $i^2=-1$. The question arises whether
$F_{p^2}$ is a field, \emph{i.e.}, one can define all the
arithmetic operations except division by zero. The definition
of addition, subtraction and multiplication in $F_{p^2}$ is
obvious and, by analogy with the field of complex numbers, one
could define division as
$1/(a+bi)\,=a/(a^2+b^2)\,-ib/(a^2+b^2)$ if $a$ and $b$ are not
equal to zero simultaneously. This definition can be meaningful
only if $a^2+b^2\neq 0$ in $F_p$. If $a$ and $b$ are not
simultaneously equal to zero, this condition can obviously be
reformulated such that $-1$ should not be a square in $F_p$ (or
in terminology of number theory it should not be a quadratic
residue). We will not consider the case $p=2$ and then $p$ is
necessarily odd. Then we have two possibilities: the value of
$p\,(mod \,4)$ is either 1 or 3. The well known result of
number theory is that -1 is a quadratic residue only in the
former case and a quadratic non-residue in the latter one,
which implies that the above construction of the field
$F_{p^2}$ is correct only if $p=3\,\,(mod \,4)$.

The main idea of establishing the correspondence between GFQT
and standard theory is as follows (see References
\cite{lev2,lev22,hep} for a detailed discussion). The first step is
to notice that the elements of $F_p$ can be written not only as
$0,1,...p-1$ but also as $0,\pm 1,...,\pm (p-1)/2$. Such
elements of $F_p$ are called minimal residues \cite{VDW,Ireland,Davenport}. 
Since the field $F_p$ is cyclic, it is convenient to visually depict
its elements by the points of a circumference of the radius
$p/2\pi$ on the plane $(x,y)$ such that the distance between
neighboring elements of the field is equal to unity and the
elements 0, 1, 2,... are situated on the circumference
counterclockwise. At the same time we depict the elements of
$Z$ as usual, such that each element $z\in Z$ is depicted by a
point with the coordinates $(z,0)$. In Fig. 1 a part of the
circumference near the origin is depicted.

\begin{figure}[!ht]
\centerline{\scalebox{1.1}{\includegraphics{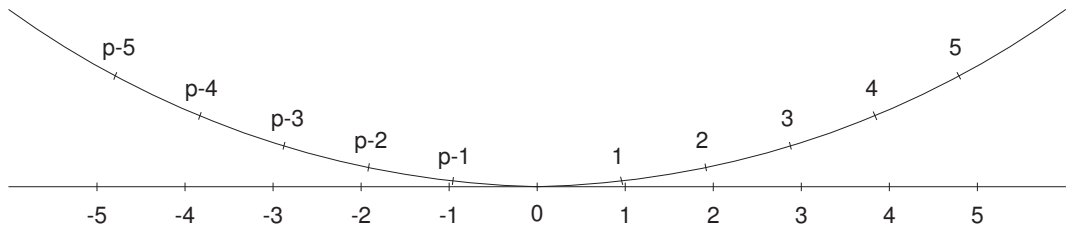}}}
\caption{
  Relation between $F_p$ and the ring of integers.
}
\label{Fig.1}
\end{figure}

Let $f$ be a map from $F_p$ to $Z$ such that $f(a)$ has the
same notation in $Z$ as its minimal residue in $F_p$. Then for
elements $a,b\in F_p$ such that $|f(a)|,|f(b)|\ll p$, addition,
subtraction and multiplication in $F_p$ and $Z$ are the same,
\emph{i.e.}, $f(a\pm b)=f(a)\pm f(b)$ and $f(ab)=f(a)f(b)$.

The second step is to establish a correspondence between
Hilbert spaces in standard theory and spaces over a Galois
field in GFQT. We first note that the Hilbert space $H$
contains a big redundancy of elements and we do not need to
know all of them. Since a set of finite linear combinations of
basis elements with rational coefficients is dense in $H$, with
any desired accuracy we can approximate each element $\tilde x$
from $H$ by a finite linear combination $ \tilde x =\tilde c_1
\tilde e_1+\tilde c_2 \tilde e_2+...\tilde c_n\tilde e_n $
where $(\tilde c_1,\tilde c_2,...\tilde c_n)$ are rational
complex numbers. In turn, the set of such elements is redundant
too. We can use the fact that Hilbert spaces in quantum theory
are projective: $\psi$ and $c\psi$ represent the same physical
state, which reflects the fact that not the probability itself
but the relative probabilities of different measurement
outcomes have a physical meaning. Therefore we can multiply
both parts of the above equality by a common denominator of the
numbers $(\tilde c_1,\tilde c_2,...\tilde c_n)$. As a result,
we can always assume that $\tilde c_j=\tilde a_j +i\tilde b_j$
where $\tilde a_j$ and $\tilde b_j$ are integers.

Consider now a space over $F_{p^2}$ and let $x =c_1 e_1+c_2
e_2+...c_n e_n$ be a decomposition of a state $x$ over a basis
$(e_1,e_2...)$ in this space. We can formally define a scalar
product $(e_j,e_k)$ such that $f((e_j,e_k))=(\tilde e_j,\tilde
e_k)$. Then the correspondence between the states $x$ and
${\tilde x}$ can be defined such that $c_j=a_j+ib_j$
$(j=1,2...)$, $f(a_j)=\tilde a_j$ and $f(b_j)=\tilde b_j$. If
the numbers in question are much less than $p$ then the
standard description and that based on GFQT give close
experimental predictions. At the same time, in GFQT a
probabilistic interpretation is not universal and is valid only
when the numbers in question are much less than $p$.

The above discussion has a well known historical analogy. For
many years people believed that our Earth was flat and
infinite, and only after a long period of time they realized
that it was finite and had a curvature. It is difficult to
notice the curvature when we deal only with distances much less
than the radius of the curvature $R$. Analogously one might
think that the set of numbers describing physics has a
curvature defined by a very large number $p$ but we do not
notice it when we deal only with numbers much less than $p$.

Since we treat GFQT as a more general theory than standard one,
it is desirable not to postulate that GFQT is based on
$F_{p^2}$ (with $p=3\,\,(mod \,4)$) because standard theory is
based on complex numbers but vice versa, explain the fact that
standard theory is based on complex numbers since GFQT is based
on $F_{p^2}$. Hence, one should find a motivation for the
choice of $F_{p^2}$ in GFQT. A possible motivation is discussed
in References \cite{hep,complex} and in Section \ref{S7} of the
present paper.

In standard approach to symmetries in quantum theory, the
symmetry group is a group of motions of a classical spacetime
background. As noted above, in quantum theory the spacetime
background does not have a physical meaning. So a question
arises whether there exists an alternative for such an
approach. As already noted, in standard approach, the spacetime
background and Lagrangian are only auxiliary tools for
constructing Hilbert spaces and operators. For calculating
observables one needs not representation operators of the
symmetry group but representation operators of its Lie algebra,
e.g., the Hamiltonian. The representation operators of the group
are needed only if it is necessary to calculate macroscopic
transformations, e.g., spacetime transformations. In the
approximation when classical time and space are good
approximate parameters, the Hamiltonian and momentum operators
can be interpreted as ones associated with the corresponding
translations, but nothing guarantees that this interpretation
is always valid (e.g., at the very early stage of the Universe).
One might think that this observation is not very significant,
since typically symmetry groups are Lie groups and for them in
many cases there exits a one-to-one correspondence between
representations of the Lie group and its Lie algebra. However,
in Galois fields there is no notion of infinitesimal
transformations and hence there is no notion of Lie group over
a Galois field associated with a given Lie algebra over a
Galois field.

Each system is described by a set of independent operators and
they somehow commute with each other. By definition, the rules
how they commute define a Lie algebra which is treated as a
symmetry algebra. Such a definition of symmetry on quantum
level is in the spirit of Dirac's paper \cite{Dir}. We believe
that for understanding this Dirac's idea the following example
might be useful. If we define how the energy should be measured
(e.g., the energy of bound states, kinetic energy \emph{etc.}),
we have a full knowledge about the Hamiltonian of our system.
In particular, we know how the Hamiltonian should commute with
other operators. In standard theory the Hamiltonian is also
interpreted as an operator responsible for evolution in time,
which is considered as a classical macroscopic parameter.
According to principles of quantum theory, self-adjoint
operators in Hilbert spaces represent observables but there is
no requirement that parameters defining a family of unitary
transformations generated by a self-adjoint operator are
eigenvalues of another self-adjoint operator. A well known
example from standard quantum mechanics is that if $P_x$ is the
$x$ component of the momentum operator then the family of
unitary transformations generated by $P_x$ is
$exp(iP_xx/\hbar)$ where $x\in (-\infty,\infty)$ and such
parameters can be identified with the spectrum of the position
operator. At the same time, the family of unitary
transformations generated by the Hamiltonian $H$ is
$exp(-iHt/\hbar)$ where $t\in (-\infty,\infty)$ and those
parameters cannot be identified with a spectrum of a
self-adjoint operator on the Hilbert space of our system. In
the relativistic case the parameters $x$ can be formally
identified with the spectrum of the Newton-Wigner position
operator \cite{NW} but it is well known that this operator does
not have all the required properties for the position operator.
So, although the operators $exp(iP_xx/\hbar)$ and
$exp(-iHt/\hbar)$ are well defined in standard theory, their
physical interpretation as translations in space and time is
not always valid.

Let us now discuss how one should define the notion of
elementary particles. Although particles are observables and
fields are not, in the spirit of quantum field theory (QFT),
fields are more fundamental than particles, and a possible
definition is as follows \cite{Wein1}: {\it It is simply a
particle whose field appears in the Lagrangian. It does not
matter if it's stable, unstable, heavy, light---if its field
appears in the Lagrangian then it's elementary, otherwise it's
composite.} Another approach has been developed by Wigner in
his investigations of unitary irreducible representations (IRs)
of the Poincare group \cite{Wigner}. In view of this approach,
one might postulate that a particle is elementary if the set of
its wave functions is the space of a unitary IR of the symmetry
group or Lie algebra in the given theory. In standard theory
the Lie algebras are usually real and one considers their
representations in complex Hilbert spaces.

In view of the above remarks, and by analogy with standard
quantum theory one might define the elementary particle in GFQT
as follows. Let ${\cal A}$ be a Lie algebra over $F_p$ which is
treated as a symmetry algebra. A particle is elementary if the
set of its states forms an IR of
${\cal A}$ in $F(p^n)$. Representations of Lie algebras in
spaces with nonzero characteristic are called modular
representations. There exists a well developed theory of such
representations. One of the well known results is the
Zassenhaus theorem \cite{Zass} that any modular IR is finite
dimensional. In Section \ref{S4} we propose another definition
of elementary particle.

As argued in References \cite{lev2,lev22,hep}, standard theories
based on de Sitter (dS) algebra so(1,4) or anti de Sitter (AdS)
algebra so(2,3) can be generalized to theories based on a
Galois field while there are problems with the generalization
of the theory based on Poincare algebra. The reasons are the
following. It is clear that in theories based on Galois fields
there can be no dimensional quantities and all physical
quantities are discrete. In standard dS or AdS invariant
theories all physical quantities are dimensionless and discrete
in units $\hbar/2=c=1$ while in Poincare invariant theory the
energy and momentum necessarily have a continuous spectrum.
From the formal point of view, the representation operators of
the Poincare algebra can also  be chosen dimensionless, e.g., in
Planck units. In Poincare invariant theories over a Galois
field one has to choose a quantum of length. If this quantum is
the Planck distance then the quantum of mass will be the Planck
mass, which is much greater than the masses of elementary
particles.

The existing astronomical data (see, e.g., Reference
\cite{Perlmutter,Melchiorri}) indicate that the cosmological constant is
small and positive. This is an argument in favor of so(1,4)
\emph{vs.} so(2,3). On the other hand, in QFT and its
generalizations (string theory, M-theory \emph{etc.}) a theory
based on so(1,4) encounters serious difficulties and the choice
of so(2,3) is preferable (see e.g., Reference \cite{Witten}). IRs
of the so(2,3) algebra have much in common with IRs of Poincare
algebra. In particular, in IRs of the so(2,3) algebra the AdS
Hamiltonian is either strictly positive or strictly negative
and a supersymmetric generalization is possible. In standard
theory, representations of the so(2,3) and so(1,4) algebras
differ only in a way how Hermiticity conditions are imposed.
Since in GFQT the notions of probability and Hermiticity are
only approximate, modular representations of those algebras
differ only in a way how we establish a correspondence with
standard theory when $p$ is large. For these reasons in the
present paper for illustration of what happens when complex
numbers are replaced by a Galois field we assume that ${\cal
A}$ is the modular analog of the algebra so(2,3).

It is well known \cite{VDW,Ireland,Davenport} that the field $F_{p^n}$ has $n-1$
nontrivial automorphisms. Therefore, if $n$ is arbitrary, a
formal scalar product and Hermiticity can be defined in
different ways. We do not assume from the beginning that $n=2$
and $p=3\,\,(mod \,4)$. Our results do not depend on the
explicit choice of the scalar product and ${\bar z}$ is used to
denote an element obtained from $z\in F_{p^n}$ by an
automorphism compatible with the scalar product in question.

The paper is organized as follows. In Sections
\ref{S2}-\ref{Singletons} we construct modular IRs describing
elementary particles in GFQT, their quantization and physical
meaning are discussed in Sections \ref{Matrix}-\ref{S6}, the
spin-statistics theorem is discussed in Section \ref{S7} and a
supersymmetric generalization is discussed in Section \ref{SS}.
Although some results require extensive calculations, they
involve only finite sums in Galois fields. For this reason all
the results can be reproduced even by readers who previously
did not have practice in calculations with Galois fields. A
more detailed description of calculations can be found in
Reference \cite{hep}.

\section{Modular IRs of the sp(2) Algebra}
\label{S2}

The key role in constructing modular IRs of the so(2,3) algebra
is played by modular IRs of the sp(2) subalgebra. They are
described by a set of operators $(a',a",h)$ satisfying the
commutation relations
\begin{equation}
[h,a']=-2a'\quad [h,a"]=2a"\quad [a',a"]=h
\label{2}
\end{equation}
The  Casimir operator of the second order for
the algebra (\ref{2}) has the form
\begin{equation}
K=h^2-2h-4a"a'=h^2+2h-4a'a"
\label{3}
\end{equation}

We first consider representations with the
vector $e_0$ such that
\begin{equation}
a'e_0=0,\quad he_0=q_0e_0
\label{4}
\end{equation}
where $q_0\in F_p$ and $f(q_0) > 0$. Recall that we consider
the representation in a linear space over $F_{p^k}$ where $k$
is a natural number (see the discussion in Section \ref{S1}).
Denote $e_n =(a")^ne_0$. Then it follows from Equations
(\ref{3}) and (\ref{4}), that \begin{equation}
he_n=(q_0+2n)e_n,\quad Ke_n=q_0(q_0-2)e_n \label{5}
\end{equation}
\begin{equation}
a'a"e_n=(n+1)(q_0+n)e_n
\label{6}
\end{equation}

One can consider analogous representations in standard theory.
Then $q_0$ is a positive real number, $n=0,1,2,...$ and the
elements $e_n$ form a basis of the IR. In this case $e_0$ is a
vector with a minimum eigenvalue of the operator $h$ (minimum
weight) and there are no vectors with the maximum weight. The
operator $h$ is positive definite and bounded below by the
quantity $q_0$. For these reasons the above modular IRs can be
treated as modular analogs of such standard IRs that $h$ is
positive definite.

Analogously, one can construct modular IRs starting from the element $e_0'$ such that
\begin{equation}
a"e_0'=0,\quad he_0'=-q_0e_0'
\label{7}
\end{equation}
and the elements $e_n'$ can be defined as $e_n'=(a')^ne_0'$.
Such modular IRs are analogs of standard IRs where $h$ is
negative definite. However, in the modular case Equations
(\ref{4}) and (\ref{7}) define the same IRs. This is clear from
the following consideration.

The set $(e_0,e_1,...e_N)$ will be a basis of IR if $a"e_i\neq
0$ for $i<N$ and $a"e_N=0$. These conditions must be compatible
with $a'a"e_N=0$. Therefore, as follows from Equation
(\ref{6}), $N$ is defined by the condition $q_0+N=0$ in $F_p$.
As a result, if $q_0$ is one of the numbers $1,...p-1$ then
$N=p-q_0$ and the dimension of IR equals $p-q_0+1$ (in
agreement with the Zassenhaus theorem \cite{Zass}). It is easy
to see that $e_N$ satisfies Equation (\ref{7}) and therefore it
can be identified with $e_0'$.

Let us forget for a moment that the eigenvalues of the operator
$h$ belong to $F_p$ and will treat them as integers. Then, as
follows from Equation (\ref{5}), the eigenvalues are
$$q_0,q_0+2,...,2p-2-q_0, 2p-q_0.$$
Therefore, if $f(q_0)>0$ and $f(q_0)\ll p$,
the maximum value of $q_0$ is equal to
$2p-q_0$, \emph{i.e.}, it is of order $2p$.

\section{Modular IRs of the so(2,3) Algebra}
\label{S3}

Standard IRs of the so(2,3) algebra relevant for describing
elementary particles have been considered by several authors.
The description in this section is a combination of two elegant
ones given in Reference \cite{Evans} for standard IRs and
Reference \cite{Braden} for modular IRs. In standard theory,
the commutation relations between the representation operators
in units $\hbar/2=c=1$ are given by
\begin{equation}
[M^{ab},M^{cd}]=-2i (g^{ac}M^{bd}+g^{bd}M^{cd}-
g^{ad}M^{bc}-g^{bc}M^{ad})
\label{8}
\end{equation}
where $a,b,c,d$ take the values 0,1,2,3,5 and the operators
$M^{ab}$ are antisymmetric. The diagonal metric tensor has the
components $g^{00}=g^{55}=-g^{11}=-g^{22}=-g^{33}=1$. In these
units the spin of fermions is odd, and the spin of bosons is
even. If $s$ is the particle spin then the corresponding IR of
the su(2) algebra has the dimension $s+1$.

Note that our definition of the AdS symmetry on quantum level
does not involve the cosmological constant at all. It appears
only if one is interested in interpreting results in terms of
the AdS spacetime or in the Poincare limit. Since all the
operators $M^{ab}$ are dimensionless in units $\hbar/2=c=1$,
the de Sitter invariant quantum theories can be formulated only
in terms of dimensionless variables. As noted in Section
\ref{S1}, this is a necessary requirement for a theory, which
is supposed to have a physical generalization to the case of
Galois fields. At the same time, since Poincare invariant
theories do not have such generalizations, one might expect
that quantities which are dimensionful in units $\hbar/2=c=1$
are not fundamental. This is in the spirit of Mirmovich's
hypothesis \cite{Mirmovich} that only quantities having the
dimension of the angular momentum can be fundamental. 

If a modular IR is considered in a linear space over $F_{p^2}$
with $p=3\,\, (mod\,\, 4)$ then Equation (\ref{8}) is also
valid. However, as noted in Section \ref{S1}, we consider
modular IRs in linear spaces over $F_{p^k}$ where $k$ is
arbitrary. In this case it is convenient to work with another
set of ten operators. Let $(a_j',a_j",h_j)$ $(j=1,2)$ be two
independent sets of operators satisfying the commutation
relations for the sp(2) algebra
\begin{equation}
[h_j,a_j']=-2a_j'\quad [h_j,a_j"]=2a_j"\quad [a_j',a_j"]=h_j
\label{9}
\end{equation}
The sets are independent in the sense that for different $j$
they mutually commute with each other. We denote additional
four operators as $b', b",L_+,L_-$. The operators
$L_3=h_1-h_2,L_+,L_-$ satisfy the commutation relations of the
su(2) algebra
\begin{equation}
[L_3,L_+]=2L_+\quad [L_3,L_-]=-2L_-\quad [L_+,L_-]=L_3
\label{10}
\end{equation}
while the other commutation relations are as follows
\begin{eqnarray}
&[a_1',b']=[a_2',b']=[a_1",b"]=[a_2",b"]=
[a_1',L_-]=[a_1",L_+]=[a_2',L_+]=\nonumber\\
&[a_2",L_-]=0\quad [h_j,b']=-b'\quad [h_j,b"]=b"\quad
[h_1,L_{\pm}]=\pm L_{\pm}\quad [h_2,L_{\pm}]=\mp L_{\pm}\nonumber\\
&[b',b"]=h_1+h_2\quad
[b',L_-]=2a_1'\quad [b',L_+]=2a_2'\quad [b",L_-]=-2a_2"\nonumber\\
&[b",L_+]=-2a_1"\quad [a_1',b"]=[b',a_2"]=L_-
\quad [a_2',b"]=[b',a_1"]=L_+ \nonumber\\
&[a_1',L_+]=[a_2',L_-]=b'\quad [a_2",L_+]=[a_1",L_-]=-b"
\label{11}
\end{eqnarray}
At first glance these relations might seem rather chaotic but
in fact they are very natural in the Weyl basis of the so(2,3)
algebra.

In spaces over $F_{p^2}$ with $p=3\,\, (mod\,\, 4)$ the
relation between the above sets of ten operators is
\begin{eqnarray}
&M_{10}=i(a_1"-a_1'-a_2"+a_2')\quad M_{15}=a_2"+a_2'-a_1"-a_1'\nonumber\\
&M_{20}=a_1"+a_2"+a_1'+a_2' \quad M_{25}=i(a_1"+a_2"-a_1'-a_2') \nonumber\\
&M_{12}=L_3\quad M_{23}=L_++L_-\quad M_{31}=-i(L_+-L_-)\nonumber\\
&M_{05}=h_1+h_2\quad M_{35}=b'+b"\quad M_{30}=-i(b"-b')
\label{12}
\end{eqnarray}
and therefore the sets are equivalent. However, the relations
(\ref{9}-\ref{11}) are more general since they can be used when
the representation space is a space over $F_{p^k}$ with an
arbitrary $k$.

We use the basis in which the operators $(h_j,K_j)$ $(j=1,2)$
are diagonal. Here $K_j$ is the Casimir operator (\ref{3}) for
algebra $(a_j',a_j",h_j)$. For constructing IRs we need
operators relating different representations of the
sp(2)$\times$sp(2) algebra. By analogy with References
\cite{Evans,Braden}, one of the possible choices is as follows
\begin{eqnarray}
&A^{++}=b"(h_1-1)(h_2-1)-a_1"L_-(h_2-1)-a_2"L_+(h_1-1)
+a_1"a_2"b'\nonumber\\
&A^{+-}=L_+(h_1-1)-a_1"b'\quad
A^{-+}=L_-(h_2-1)-a_2"b'\quad A^{--}=b'
\label{13}
\end{eqnarray}
We consider the action of these operators only on the space of
minimal sp(2)$\times$sp(2) vectors, \emph{i.e.}, such vectors
$x$ that $a_j'x=0$ for $j=1,2$, and $x$ is the eigenvector of
the operators $h_j$. If $x$ is a minimal vector such that
$h_jx=\alpha_jx$ then $A^{++}x$ is the minimal eigenvector of
the operators $h_j$ with the eigenvalues $\alpha_j+1$,
$A^{+-}x$ - with the eigenvalues $(\alpha_1+1,\alpha_2-1)$,
$A^{-+}x$ - with the eigenvalues $(\alpha_1-1,\alpha_2+1)$, and
$A^{--}x$ - with the eigenvalues $\alpha_j-1$.

By analogy with References \cite{Evans,Braden}, we require
the existence of the vector $e_0$ satisfying the conditions
\begin{eqnarray}
&a_j'e_0=b'e_0=L_+e_0=0\quad h_je_0=q_je_0\quad (j=1,2)
\label{15}
\end{eqnarray}
where $q_j\in F_p$, $f(q_j)>0$ for $j=1,2$ and $f(q_1-q_2)\geq
0$. It is well known (see e.g., Reference \cite{hep}) that
$M^{05}=h_1+h_2$ is the AdS analog of the energy operator. As
follows from Equations (\ref{9}) and (\ref{11}), the operators
$(a_1',a_2',b')$ reduce the AdS energy by two units. Therefore
$e_0$ is an analog the state with the minimum energy which can
be called the rest state, and the spin in our units is equal to
the maximum value of the operator $L_3=h_1-h_2$ in that state.
For these reasons we use $s$ to denote $q_1-q_2$ and $m$ to
denote $q_1+q_2$. In standard classification \cite{Evans}, the
massive case is characterized by the condition $q_2>1$ and the
massless one---by the condition $q_2=1$. There also exist two
exceptional IRs discovered by Dirac \cite{DiracS} (Dirac
singletons). They are characterized by the conditions $m=1,\,\,
s=0$ and $m=2,\,\, s=1$. In this section we will consider the
massive case while the singleton and massless cases will be
considered in the next section.

As follows from the above remarks, the elements
\begin{equation}
e_{nk}=(A^{++})^n(A^{-+})^ke_0
\label{16}
\end{equation}
represent the minimal sp(2)$\times$sp(2) vectors with the
eigenvalues of the operators $h_1$ and $h_2$ equal to
$Q_1(n,k)=q_1+n-k$ and $Q_2(n,k)=q_2+n+k$, respectively. It can
be shown by a direct calculation that
\begin{equation}
A^{--}A^{++}e_{nk}=(n+1)(m+n-2)(q_1+n)(q_2+n-1)e_{nk}
\label{17}
\end{equation}
\begin{equation}
A^{+-}A^{-+}e_{nk}=(k+1)(s-k)(q_1-k-2)(q_2+k-1)e_{nk}
\label{18}
\end{equation}

As follows from these expressions, in the massive case $k$ can
assume only the values $0,1,...s$ and in standard theory
$n=0,1,...\infty$. However, in the modular case
$n=0,1,...n_{max}$ where $n_{max}$ is the first number for
which the r.h.s. of Equations (\ref{17}) becomes zero in $F_p$.
Therefore $n_{max}=p+2-m$.

The full basis of the representation space can be chosen in the
form
\begin{equation}
e(n_1n_2nk)=(a_1")^{n_1}(a_2")^{n_2}e_{nk}
\label{19}
\end{equation}
In standard theory $n_1$ and $n_2$ can be any
natural numbers. However, as follows from the
results of the preceding section, Equation (\ref{9}) and
the properties of the $A$ operators,
\begin{eqnarray}
&n_1=0,1,...N_1(n,k)\quad n_2=0,1,...N_2(n,k)\nonumber\\
&N_1(n,k)=p-q_1-n+k\quad N_2(n,k)=p-q_2-n-k
\label{20}
\end{eqnarray}
As a consequence, the representation is finite dimensional in
agreement with the Zassenhaus theorem \cite{Zass} (moreover, it
is finite since any Galois field is finite).

Let us assume additionally that the representation space is
supplied by a scalar product (see Section \ref{S1}). The
element $e_0$ can always be chosen such that $(e_0,e_0)=1$.
Suppose that the representation operators satisfy the
Hermiticity conditions $L_+^*=L_-$, $a_j^{'*}=a_j"$,
$b^{'*}=b"$ and $h_j^*=h_j$. Then, as follows from Equation
(\ref{12}), in a special case when the representation space is
a space over $F_{p^2}$ with $p=3\,\, (mod\,\, 4)$, the
operators $M^{ab}$ are Hermitian as it should be. By using
Equations (\ref{9}-\ref{18}), one can show by a direct
calculation that the elements $e(n_1n_2nk)$ are mutually
orthogonal while the quantity
\begin{equation}
Norm(n_1n_2nk)=(e(n_1n_2nk),e(n_1n_2nk))
\label{23}
\end{equation}
can be represented as
\begin{equation}
Norm(n_1n_2nk)=F(n_1n_2nk)G(nk)
\label{24}
\end{equation}
where
\begin{eqnarray}
&F(n_1n_2nk)= n_1!(Q_1(n,k)+n_1-1)!n_2!(Q_2(n,k)+n_2-1)!\nonumber\\
&G(nk)=\{(q_2+k-2)!n!(m+n-3)!(q_1+n-1)!(q_2+n-2)!k!s!\}\nonumber\\
&\{(q_1-k-2)![(q_2-2)!]^3(q_1-1)!(m-3)!(s-k)!\nonumber\\
&[Q_1(n,k)-1][Q_2(n,k)-1]\}^{-1}
\label{25}
\end{eqnarray}

In standard Poincare and AdS theories there also exist IRs with
negative energies. They can be constructed by analogy with
positive energy IRs. Instead of Equation (\ref{15}) one can
require the existence of the vector $e_0'$ such that
\begin{eqnarray}
&a_j"e_0'=b"e_0'=L_-e_0'=0\quad h_je_0'=-q_je_0'\quad (e_0',e_0')\neq 0\quad (j=1,2)
\label{26}
\end{eqnarray}
where the quantities $q_1,q_2$ are the same as for positive
energy IRs. It is obvious that positive and negative energy IRs
are fully independent since the spectrum of the operator
$M^{05}$ for such IRs is positive and negative, respectively.
However, {\it the modular analog of a positive energy IR
characterized by $q_1,q_2$ in Equation (\ref{15}), and the
modular analog of a negative energy IR characterized by the
same values of $q_1,q_2$ in Equation (\ref{26}) represent the
same modular IR.} This is the crucial difference between
standard quantum theory and GFQT, and a proof is given below.

\begin{sloppypar}
Let $e_0$ be a vector satisfying Equation (\ref{15}). Denote
$N_1=p-q_1$ and $N_2=p-q_2$. Our goal is to prove that the
vector $x=(a_1")^{N_1}(a_2")^{N_2}e_0$ satisfies the conditions
(\ref{26}), \emph{i.e.}, $x$ can be identified with $e_0'$.
\end{sloppypar}

As follows from the definition of $N_1,N_2$, the vector $x$ is
the eigenvector of the operators $h_1$ and $h_2$ with the
eigenvalues $-q_1$ and $-q_2$, respectively, and, in addition,
it satisfies the conditions $a_1"x=a_2"x=0$. Let us prove that
$b"x=0$. Since $b"$ commutes with the $a_j"$, we can write
$b"x$ in the form
\begin{equation}
b"x = (a_1")^{N_1}(a_2")^{N_2}b"e_0
\label{27}
\end{equation}
As follows from Equations (\ref{11}) and (\ref{15}),
$a_2'b"e_0=L_+e_0=0$ and $b"e_0$ is the eigenvector of the
operator $h_2$ with the eigenvalue $q_2+1$. Therefore, $b"e_0$
is the minimal vector of the sp(2) IR which has the dimension
$p-q_2=N_2$. Therefore $(a_2")^{N_2}b"e_0=0$ and $b"x=0$.

The next stage of the proof is to show that $L_-x=0$.
As follows from Equation (\ref{11}) and the definition of
$x$,
\begin{equation}
L_-x = (a_1")^{N_1}(a_2")^{N_2}L_-e_0-
N_1(a_1")^{N_1-1}(a_2")^{N_2}b"e_0
\label{28}
\end{equation}
We have already shown that $(a_2")^{N_2}b"e_0=0$, and therefore
it is sufficient to prove that the first term in the r.h.s. of
Equation (\ref{28}) is equal to zero. As follows from Equations
(\ref{11}) and (\ref{15}), $a_2'L_-e_0=b'e_0=0$, and $L_-e_0$
is the eigenvector of the operator $h_2$ with the eigenvalue
$q_2+1$. Therefore $(a_2")^{N_2}L_-e_0=0$ and the proof is
completed.

Let us assume for a moment that the eigenvalues of the
operators $h_1$ and $h_2$ should be treated not as elements of
$F_p$ but as integers. Then, as follows from the consideration
in the preceding section, if $f(q_j)\ll p$ (j=1,2) then one
modular IR of the so(2,3) algebra corresponds to a standard
positive energy IR in the region where the energy is positive
and much less than $p$. At the same time, it corresponds to an
IR with the negative energy in the region where the AdS energy
is close to $4p$ but less than $4p$.

\section{Massless Particles and Dirac Singletons}
\label{Singletons}

Those cases can be considered by analogy with the massive one.
The case of Dirac singletons is especially simple. As follows
from Equations (\ref{17}) and (\ref{18}), if $m=1,\,\, s=0$
then the only possible value of $k$ is $k=0$ and the only
possible values of $n$ are $n=0,1$ while if $m=2,\,\, s=1$ then
the only possible values of $k$ are $k=0,1$ and the only
possible value of $n$ is $n=0$. This result does not depend on
the value of $p$ and therefore it is valid in both, standard
theory and GFQT. In this case the only difference between
standard and modular cases is that in the former
$n_1,n_2=0,1,...\infty$ while in the latter the quantities
$n_1,n_2$ are in the range defined by Equation (20).

The singleton IRs are indeed exceptional since the value of $n$
in them does not exceed 1 and therefore the impression is that
singletons are two-dimensional objects, not three-dimensional
ones as usual particles. However, the singleton IRs have been
obtained in the so(2,3) theory without reducing the algebra.
Dirac has entitled his paper \cite{DiracS} "A Remarkable
Representation of the 3 + 2 de Sitter Group". Below we argue
that in GFQT the singleton IRs are even more remarkable than in
standard theory.

If $m=1,\,\, s=0$ then $q_1=q_2=1/2$. In GFQT these relations
should be treated as $q_1=q_2=(p+1)/2$. Analogously, if
$m=2,\,\, s=1$ then $q_1=3/2,\,\, q_2=1/2$ and in GFQT
$q_1=(p+3)/2,\,\, q_2=(p+1)/2$. Therefore when the values of
$n_1$ and $n_2$ are small, the values of $h_1$ and $h_2$ are
extremely large since they are of order of $p/2$. As follows
from the results of Sections \ref{S2} and \ref{S3}. those
values are much less than $p$ only when $n_1$ and $n_2$ are of
order $p/4$. This might be an indication why singletons are not
observable: because there is no region when all the quantum
numbers are much less than $p$. At the end of this section we
will discuss relations between singleton and massless IRs.

Consider now the massless case. We will follow our derivation
in Reference \cite{tmf}. When $q_2=1$, it is more convenient to
deal not with the $A$-operators defined in Equation (\ref{13})
but with the $B$-operators defined as
\begin{eqnarray}
&B^{++}=b"-a_1"L_-(h_1-1)^{-1}-a_2"L_+(h_2-1)^{-1}+a_1"a_2"b'[(h_1-1)(h_2-1)]^{-1}\nonumber\\
&B^{+-}=L_+-a_1"b'(h_1-1)^{-1}\quad B^{-+}=L_--a_2"b'(h_2-1)^{-1}\quad B^{--}=b'
\label{M12}
\end{eqnarray}
If $e_0$ is defined as in Equation (\ref{15}), then by, analogy
with the massive case, we can define the vectors $e_{nk}$ as
\begin{equation}
e_{nk}=(B^{++})^n(B^{-+})^ke_0
\label{M16}
\end{equation}
but a problem arises how to define the action of the operators
$B^{++}$ and $B^{-+}$ on $e_0$ which is the eigenvector of the
operator $h_2$ with the eigenvalue $q_2=1$. A possible way to
resolve ambiguities 0/0 in matrix elements is to write $q_2$ in
the form $q_2=1+\epsilon$ and take the limit
$\epsilon\rightarrow 0$ at the final stage of computations.
This confirms a well known fact that analytical methods can be
very useful in problems involving only integers. It is also
possible to justify the results by using only integers (or
rather elements of the Galois field in question), but we will
not dwell on this .

By using the above prescription, we require that
\begin{eqnarray}
B^{++}e_0=[b"-a_1"L_-(h_1-1)^{-1}]e_0\quad B^{-+}e_0=L_-e_0
\label{M14}
\end{eqnarray}
if $s\neq 0$ (and thus $h_1\neq 1$), and
\begin{equation}
B^{++}e_0=b"e_0\quad B^{+-}e_0=B^{-+}e_0=0
\label{M15}
\end{equation}
if $s=0$. One can directly verify that, as follows from
Equations (\ref{9}-\ref{11})
\begin{eqnarray}
&B^{-+}B^{++}(h_1-1)=B^{++}B^{-+}(h_1-2)\quad B^{+-}B^{++}(h_2-1)=B^{++}B^{+-}(h_2-2)
\label{M17}
\end{eqnarray}
and, in addition, as follows from Equation (\ref{15})
\begin{equation}
B^{--}e_{nk}=a(n,k)e_{n-1,k}\quad
B^{+-}e_{nk}=b(n,k)e_{n,k-1}
\label{M18}
\end{equation}
where
\begin{eqnarray}
&a(n,k)=\frac{n(n+s-1)(n+s)(n-1)}{(n+s-k-1)(n+k-1)}\quad b(n,k)=\frac{k(s+1-k)(k-1)}{n+k-1}
\label{M19}
\end{eqnarray}
As follows from these expressions, the elements $e_{nk}$ form a
basis in the space of minimal sp(2)$\times$sp(2) vectors, and
our next goal is to determine the range of the numbers $n$ and
$k$.

Consider first the quantity $b(0,k)=k(s+1-k)$ and let $k_{max}$
be the maximum value of $k$. For consistency we should require
that if $k_{max}\neq 0$ then $k=k_{max}$ is the greatest value
of $k$ such that $b(0,k) \neq 0$ for $k=1,...k_{max}$. We
conclude that $k$ can take only the values of $0,1,..s$.

Let now $n_{max}(k)$ be the maximum value of $n$ at a given
$k$. For consistency we should require that if $n_{max}(k) \neq
0$ then $n_{max}(k)$ is the greatest value of $n$ such that
$a(n,k)\neq 0$ for $n=1,...n_{max}(k)$. As follows from
Equation (\ref{M19}), $a(1,k)=0$ for $k=1,..s-1$ if such values
of $k$ exist (\emph{i.e.}, when $s\geq 2$), and $a(n,k)=n(s+n)$
if $k=0$ or $k=s$. We conclude that at $k=1,...s-1$, the
quantity $n$ can take only the value $n=0$ while at $k=0$ or
$k=s$, the possible values of $n$ are $0,1,...n_{max}$ where
$n_{max}=p-s-1$. Recall that in the preceding section we have
obtained $n_{max}=p+2-m$ for the massive case. Since $m=2q_2+s$
and $q_2>1$ in the massive case, we conclude that the values of
$n_{max}$ in the massive and massless cases are given by
different formulas.

According to Standard Model, only massless Weyl particles can
be fundamental elementary particles in Poincare invariant
theory. Therefore a problem arises whether the above results
can be treated as analogs of Weyl particles in standard and
modular versions of AdS invariant theory. Several authors
investigated dS and AdS analogs of Weyl particles proceeding
from covariant equations on the dS and AdS spaces,
respectively. For example, the authors of Reference
\cite{Ikeda} have shown that Weyl particles arise only when the
dS or AdS symmetries are broken to the Lorentz symmetry. The
results of Reference \cite{Evans} and the above results in the
modular case make it possible to treat AdS Weyl particles from
the point of view of IRs.

It is well known that Poincare invariant theory is a special
case of AdS one obtained as follows. We introduce the AdS
radius $R$ and define $P^{\mu}=M^{\mu 5}/2R$ $(\mu =0,1,2,3)$.
Then in the approximation when $R$ is very large, the operators
$M^{\mu 5}$ are very large but their ratio is finite, we obtain
Poincare invariant theory where $P^{\mu}$ are the four-momentum
operators. This procedure is called contraction and for the
first time it has been discussed in Reference \cite{IW}. Since
the mass is the lowest value of the energy in both, Poincare
and AdS invariant theories, the mass $m$ in the AdS case and
the standard Poincare mass $m'$ are related as $m/2R=m'$. The
AdS mass is dimensionless while the Poincare mass has the
dimension $length^{-1}$. Since the Poincare symmetry is a
special case of the AdS one, this fact is in agreement with the
observation in Section \ref{S1} that dimensionful quantities
cannot be fundamental. Let $l_C(m')$ be the Compton wave length
for the particle with the mass $m'$. Then one might think that,
in view of the relation $m=2R/l_C(m')$, the AdS mass shows how
many Compton wave lengths are contained in the interval
$(0,2R)$. However, such an interpretation of the AdS mass means
that we wish to interpret a fundamental quantity $m$ in terms
of our experience based on Poincare invariant theory. As
already noted, the value of $m$ does not depend on any
quantities having the dimension $length$ or $length^{-1}$ and
it is the Poincare mass which implicitly depends on $R$. Let us
assume for estimations that the value of $R$ is $10^{28}cm$.
Then even the AdS mass of the electron is of order $10^{39}$
and this might be an indication that the electron is not a true
elementary particle. Moreover, the present upper level for the
photon mass is $10^{-18}ev$ which seems to be an extremely tiny
quantity. However, the corresponding AdS mass is of order
$10^{15}$ and so even the mass which is treated as extremely
small in Poincare invariant theory might be very large in AdS
invariant theory.

Since $m=2q_2+s$, the corresponding Poincare mass will be zero
when $R\to\infty$ not only when $q_2=1$ but when $q_2$ is any
finite number. So a question arises why only the case $q_2=1$
is treated as massless. In Poincare invariant theory, Weyl
particles are characterized not only by the condition that
their mass is zero but also by the condition that they have a
definite helicity. In standard case the minimum value of the
AdS energy for massless IRs with positive energy is
$E_{min}=2+s$ when $n=0$. In contrast with the situation in
Poincare invariant theory, where massless particles cannot be
in the rest state, the massless particles in the AdS theory do
have rest states and, as shown above, the value of the $z$
projection of the spin in such states can be $-s,-s+2, ...s$ as
usual. However, we have shown that for any value of energy
greater than $E_{min}$, when $n\neq 0$, the spin state is
characterized only by helicity, which can take the values
either $s$ when $k=0$ or $-s$ when $k=s$, \emph{i.e.}, we have
the same result as in Poincare invariant theory. Note that in
contrast with IRs of the Poincare and dS algebras, standard IRs
describing particles in AdS invariant theory belong to the
discrete series of IRs and the energy spectrum in them is
discrete: $E=E_{min}, E_{min}+2, ...\infty$. Therefore,
strictly speaking, rest states do not have measure zero as in
Poincare and dS invariant theories. Nevertheless, the
probability that the energy is exactly $E_{min}$ is extremely
small and therefore the above results show that the case
$q_2=1$ indeed describes AdS analogs of Weyl particles.

By analogy with the massive case, one can show that the full
basis of the representation space also can be described by
Equation (\ref{19}) and that one massless modular IR is a
modular analog of both, standard massless positive and negative
energy IRs. For singleton IRs it is also possible to prove that
if a vector $e_0'$ is defined by the same formulas as in
Section \ref{S3}, it satisfies Equation (\ref{26}). However,
singleton IRs obviously cannot be treated as modular analogs of
standard positive and negative energy IRs.

In Reference \cite{FF} entitled "One Massless Particle Equals
Two Dirac Singletons", it is shown that the tensor product of
two singleton IRs is a massless IR. This follows from the following 
facts. If we take two singleton IRs then the
tensor product of the corresponding vectors $e_0$ (see Equation
(\ref{15})) satisfies Equation (\ref{15}) and is characterized
by $q_2=1$, \emph{i.e.}, precisely by the condition defining a
massless IR. The value of spin in this IR equals 0 for the
tensor product of two singletons with $m=1,\,\,s=0$, 1
(\emph{i.e.}, 1/2 in standard units) for the tensor product of
two singleton IRs with $m=1,\,\,s=0$ and $m=2,\,\,s=1$ and 2
(\emph{i.e.}, 1 in standard units) for the tensor product of two
singletons with $m=2,\,\,s=1$. Therefore the tensor product of two 
singleton IRs indeed contains a massless IR and, as a consequence
of a special nature of singleton IRs, it does not 
contain other IRs. This might be an indication that
fundamental particles are even not Weyl ones but Dirac
singletons. We believe that in GFQT the singleton IRs are even
more remarkable than in standard theory for the following
reasons. If we accept that Weyl particles are composite states
of Dirac singletons then a question arises why Weyl particles
are stable and singletons have not been observed yet although
in standard theory they are characterized by small values of
all quantum numbers. However, in GFQT at least two singleton
quantum numbers are of order $p$, \emph{i.e.}, extremely large
and this might be an explanation why they are not observable in
situations where all energies in question are much less than
$p$. We believe this is an interesting observation that when
the values of $h_1$ and $h_2$ are of order $p/2$, their sum is
small since it is calculated modulo $p$. In standard theory, if
an additive quantity for a two-particle system is not equal to
a sum of the corresponding single-particle quantities, it is
said that the particles interact. Therefore the fact that a sum
of two values of order $p/2$ is not of order $p$ but much less
than $p$ can be treated as a very strong interaction although
from the formal point of view no interaction between the
singletons has been introduced.

\section{Matrix Elements of Representation Operators}
\label{Matrix}

In what follows, we will discuss the massive case but the same
results are valid in the singleton and massless cases. The
matrix elements of the operator $A$ are defined as
\begin{equation}
Ae(n_1n_2nk)=\sum_{n_1'n_2'n'k'}
A(n_1'n_2'n'k';n_1n_2nk)e(n_1'n_2'n'k')
\label{29}
\end{equation}
where the sum is taken over all possible values of
$(n_1'n_2'n'k')$. One can explicitly calculate matrix elements
for all the representation operators and the results are as
follows.
\begin{eqnarray}
&h_1e(n_1n_2nk)=[Q_1(n,k)+2n_1]e(n_1n_2nk)\nonumber\\
& h_2e(n_1n_2nk)=[Q_2(n,k)+2n_2]e(n_1n_2nk)
\label{30}
\end{eqnarray}
\begin{eqnarray}
&a_1'e(n_1n_2nk)=n_1[Q_1(n,k)+n_1-1]e(n_1-1,n_2nk)\nonumber\\
&a_1"e(n_1n_2nk)=e(n_1+1,n_2nk)\nonumber\\
&a_2'e(n_1n_2nk)=n_2[Q_2(n,k)+n_2-1]e(n_1,n_2-1,nk)\nonumber\\
&a_2"e(n_1n_2nk)=e(n_1,n_2+1,nk)
\label{31}
\end{eqnarray}
\begin{eqnarray}
&b"e(n_1n_2nk)=\{[Q_1(n,k)-1][Q_2(n,k)-1]\}^{-1}\nonumber\\
&[k(s+1-k)(q_1-k-1)(q_2+k-2)e(n_1,n_2+1,n,k-1)+\nonumber\\
&n(m+n-3)(q_1+n-1)(q_2+n-2)e(n_1+1,n_2+1,n-1,k)+\nonumber\\
&e(n_1,n_2,n+1,k)+e(n_1+1,n_2,n,k+1)]
\label{32}
\end{eqnarray}
\begin{eqnarray}
&b'e(n_1n_2nk)=\{[Q_1(n,k)-1][Q_2(n,k)-1]\}^{-1}
[n(m+n-3)\nonumber\\
&(q_1+n-1)(q_2+n-2)(q_1+n-k+n_1-1)(q_2+n+k+n_2-1)\nonumber\\
&e(n_1n_2,n-1,k)+n_2(q_1+n-k+n_1-1)e(n_1,n_2-1,n,k+1)+\nonumber\\
&n_1(q_2+n+k+n_2-1)k(s+1-k)(q_1-k-1)(q_2+k-2)\nonumber\\
&e(n_1-1,n_2,n,k-1)+n_1n_2e(n_1-1,n_2-1,n+1,k)]
\label{33}
\end{eqnarray}
\begin{eqnarray}
&L_+e(n_1n_2nk)=\{[Q_1(n,k)-1][Q_2(n,k)-1]\}^{-1}
\{(q_2+n+k+n_2-1)\nonumber\\
&[k(s+1-k)(q_1-k-1)(q_2+k-2)e(n_1n_2n,k-1)+\nonumber\\
&n(m+n-3)(q_1+n-1)(q_2+n-2)e(n_1+1,n_2,n-1,k)]+\nonumber\\
&n_2[e(n_1,n_2-1,n+1,k)+e(n_1+1,n_2-1,n,k+1)]\}
\label{34}
\end{eqnarray}
\begin{eqnarray}
&L_-e(n_1n_2nk)=\{[Q_1(n,k)-1][Q_2(n,k)-1]\}^{-1}
\{n_1[k(s+1-k)\nonumber\\
&(q_1-k-1)(q_2+k-2)e(n_1-1,n_2n,k-1)+e(n_1-1,n_2,n+1,k)]\nonumber\\
&+(q_1+n-k+n_1-1)[e(n_1n_2n,k+1)+n(m+n-3)\nonumber\\
&(q_1+n-1)(q_2+n-2)e(n_1,n_2+1,n-1,k)]\}
\label{35}
\end{eqnarray}
We will always use a convention that $e(n_1n_2nk)$ is a null
vector if some of the numbers $(n_1n_2nk)$ are not in the range
described above.

The important difference between standard and modular IRs is
that in the latter the trace of each representation operator is
equal to zero while in the former this is obviously not the
case (for example, the energy operator is positive definite for
IRs defined by Equation (\ref{15}) and negative definite for
IRs defined by Equation (\ref{26})). For the operators
$(a_j',a_j",L_{\pm},b',b")$ the validity of this statement is
clear immediately: since they necessarily change one of the
quantum numbers $(n_1n_2nk)$, they do not contain nonzero
diagonal elements at all. The proof for the diagonal operators
$h_1$ and $h_2$ is as follows. For each IR of the sp(2) algebra
with the "minimal weight" $q_0$ and the dimension $N+1$, the
eigenvalues of the operator $h$ are $(q_0,q_0+2,...q_0+2N)$.
The sum of these eigenvalues equals zero in $F_p$ since
$q_0+N=0$ in $F_p$ (see the preceding section). Therefore we
conclude that for any representation operator $A$
\begin{equation}
\sum_{n_1n_2nk} A(n_1n_2nk,n_1n_2nk)=0
\label{36}
\end{equation}
This property is very important for investigating a new
symmetry between particles and antiparticles in the GFQT
which is discussed in the subsequent section.

\section{Quantization and AB Symmetry}
\label{S4}

Let us first recall how the Fock space is defined in standard
theory. Let $a(n_1n_2nk)$ be the operator of particle
annihilation in the state described by the vector
$e(n_1n_2nk)$. Then the adjoint operator $a(n_1n_2nk)^*$ has
the meaning of particle creation in that state. Since we do not
normalize the states $e(n_1n_2nk)$ to one, we require that the
operators $a(n_1n_2nk)$ and $a(n_1n_2nk)^*$ should satisfy
either the anticommutation relations
\begin{eqnarray}
\{a(n_1n_2nk),a(n_1'n_2'n'k')^*\}=Norm(n_1n_2nk)
\delta_{n_1n_1'}\delta_{n_2n_2'}\delta_{nn'}\delta_{kk'}
\label{37}
\end{eqnarray}
or the commutation relations
\begin{eqnarray}
[a(n_1n_2nk),a(n_1'n_2'n'k')^*]=Norm(n_1n_2nk)
\delta_{n_1n_1'}\delta_{n_2n_2'}\delta_{nn'}\delta_{kk'}
\label{38}
\end{eqnarray}

In standard theory the representation describing a particle and
its antiparticle are fully independent and therefore
quantization of antiparticles should be described by other
operators. If $b(n_1n_2nk)$ and $b(n_1n_2nk)^*$ are operators
of the antiparticle annihilation and creation in the state
$e(n_1n_2nk)$ then by analogy with Equations (\ref{37}) and
(\ref{38})
\begin{eqnarray}
\{b(n_1n_2nk),b(n_1'n_2'n'k')^*\}=Norm(n_1n_2nk)
\delta_{n_1n_1'}\delta_{n_2n_2'}\delta_{nn'}\delta_{kk'}
\label{39}
\end{eqnarray}
\begin{eqnarray}
[b(n_1n_2nk),b(n_1'n_2'n'k')^*]=Norm(n_1n_2nk)
\delta_{n_1n_1'}\delta_{n_2n_2'}\delta_{nn'}\delta_{kk'}
\label{40}
\end{eqnarray}
for anticommutation or commutation relations, respectively. In
this case it is assumed that in the case of anticommutation
relations all the operators $(a,a^*)$ anticommute with all the
operators $(b,b^*)$ while in the case of commutation relations
they commute with each other. It is also assumed that the Fock
space contains the vacuum vector $\Phi_0$ such that
\begin{equation}
a(n_1n_2nk)\Phi_0=b(n_1n_2nk)\Phi_0=0\quad
\forall\,\, n_1,n_2,n,k
\label{41}
\end{equation}

The Fock space in standard theory can now be defined as a
linear combination of all elements obtained by the action of
the operators $(a^*,b^*)$ on the vacuum vector, and the problem
of second quantization of representation operators can be
formulated as follows. Let $(A_1,A_2....A_n)$ be representation
operators describing IR of the AdS algebra. One should replace
them by operators acting in the Fock space such that the
commutation relations between their images in the Fock space
are the same as for original operators (in other words, we
should have a homomorphism of Lie algebras of operators acting
in the space of IR and in the Fock space). We can also require
that our map should be compatible with the Hermitian
conjugation in both spaces. It is easy to verify that a
possible solution satisfying all the requirements is as
follows. Taking into account the fact that the matrix elements
satisfy the proper commutation relations, the operators $A_i$
in the quantized form
\begin{eqnarray}
&A_i=\sum A_i(n_1'n_2'n'k',n_1n_2nk)
[a(n_1'n_2'n'k')^*a(n_1n_2nk)+\nonumber\\
&b(n_1'n_2'n'k')^*b(n_1n_2nk)]/Norm(n_1n_2nk)
\label{42}
\end{eqnarray}
satisfy the commutation relations (\ref{9}-\ref{11}). We will
not use special notations for operators in the Fock space since
in each case it will be clear whether the operator in question
acts in the space of IR or in the Fock space.

\begin{sloppypar}
A well known problem in standard theory is that the
quantization procedure does not define the order of the
annihilation and creation operators uniquely. For example,
another possible solution is
\begin{eqnarray}
&A_i=\mp \sum A_i(n_1'n_2'n'k',n_1n_2nk)
[a(n_1n_2nk)a(n_1'n_2'n'k')^*+\nonumber\\
&b(n_1n_2nk)b(n_1'n_2'n'k')^*]/Norm(n_1n_2nk)
\label{43}
\end{eqnarray}
for anticommutation and commutation relations, respectively.
The solutions (\ref{42}) and (\ref{43}) are different since the
energy operators $M^{05}$ in these expressions differ by an
infinite constant. In standard theory the solution (\ref{42})
is selected by imposing an additional requirement that all
operators should be written in the normal form where
annihilation operators precede creation ones. Then the vacuum
has zero energy and Equation (\ref{43}) should be rejected.
Such a requirement does not follow from the theory. Ideally
there should be a procedure which correctly defines the order
of operators from first principles.
\end{sloppypar}

In standard theory there also exist neutral particles. In that
case there is no need to have two independent sets of operators
$(a,a^*)$ and $(b,b^*)$, and Equation (\ref{42}) should be
written without the $(b,b^*)$ operators. The problem of neutral
particles in GFQT is discussed in Section \ref{S7}.

We now proceed to quantization in the modular case. The results
of Section \ref{S3} show that one modular IR corresponds to two
standard IRs with the positive and negative energies,
respectively. This indicates to a possibility that one modular
IR describes a particle and its antiparticle simultaneously.
However, we don't know yet what should be treated as a particle
and its antiparticle in the modular case. We have a description
of an object such that $(n_1n_2nk)$ is the full set of its
quantum numbers which take the values described in the
preceding section.

We now assume that $a(n_1n_2nk)$ in GFQT is the operator
describing annihilation of the object with the quantum numbers
$(n_1n_2nk)$ regardless of whether the numbers are physical or
nonphysical. Analogously $a(n_1n_2nk)^*$ describes creation of
the object with the quantum numbers $(n_1n_2nk)$. If these
operators anticommute then they satisfy Equation (\ref{37})
while if they commute then they satisfy Equation (\ref{38}).
Then, by analogy with standard case, the operators
\begin{eqnarray}
A_i=\sum A_i(n_1'n_2'n'k',n_1n_2nk)a(n_1'n_2'n'k')^*a(n_1n_2nk)/Norm(n_1n_2nk)
\label{44}
\end{eqnarray}
satisfy the commutation relations (\ref{9}-\ref{11}). In this
expression the sum is taken over all possible values of the
quantum numbers in the modular case.

In the modular case the solution can be taken not only as in
Equation (\ref{44}) but also as
\begin{eqnarray}
A_i=\mp\sum A_i(n_1'n_2'n'k',n_1n_2nk)a(n_1n_2nk)a(n_1'n_2'n'k')^*/Norm(n_1n_2nk)
\label{45}
\end{eqnarray}
for the cases of anticommutators and commutators, respectively.
However, as follows from Equations (\ref{36}-\ref{38}), the
solutions (\ref{44}) and (\ref{45}) are the same. Therefore in
the modular case there is no need to impose an artificial
requirement that all operators should be written in the normal
form.

The problem with the treatment of the $(a,a^*)$ operators is as
follows.  When the values of $(n_1n_2n)$ are much less than
$p$, the modular IR corresponds to standard positive energy IR
and therefore the $(a,a^*)$ operator can be treated as those
describing the particle annihilation and creation,
respectively. However, when the AdS energy is negative, the
operators $a(n_1n_2nk)$ and $a(n_1n_2nk)^*$ become unphysical
since they describe annihilation and creation, respectively, in
the unphysical region of negative energies.

Let us recall that at any fixed values of $n$ and $k$, the
quantities $n_1$ and $n_2$ can take only the values described
in Equation (\ref{20}) and the eigenvalues of the operators
$h_1$ and $h_2$ are given by $Q_1(n,k)+2n_1$ and
$Q_2(n,k)+2n_2$, respectively. As follows from the results of
Section \ref{S3}, the first IR of the sp(2) algebra has the
dimension $N_1(n,k)+1$ and the second IR has the dimension
$N_2(n,k)+1$. If $n_1=N_1(n,k)$ then it follows from Equation
(\ref{20}) that the first eigenvalue is equal to $-Q_1(n,k)$ in
$F_p$, and if $n_2=N_2(n,k)$ then the second eigenvalue is
equal to $-Q_2(n,k)$ in $F_p$. We use ${\tilde n}_1$ to denote
$N_1(n,k)-n_1$ and ${\tilde n}_2$ to denote $N_2(n,k)-n_2$.
Then it follows from Equation (\ref{20}) that $e({\tilde
n}_1{\tilde n}_2nk)$ is the eigenvector of the operator $h_1$
with the eigenvalue $-(Q_1(n,k)+2n_1)$ and the eigenvector of
the operator $h_2$ with the eigenvalue $-(Q_2(n,k)+2n_2)$.

Standard theory implicitly involves the idea that creation of
the antiparticle with positive energy can be treated as
annihilation of the corresponding particle with negative energy
and annihilation of the antiparticle with positive energy can
be treated as creation of the corresponding particle with
negative energy. In GFQT we can implement this idea explicitly.
Namely, we can define the operators $b(n_1n_2nk)$ and
$b(n_1n_2nk)^*$ in such a way that they will replace the
$(a,a^*)$ operators if the quantum numbers are unphysical. In
addition, if the values of $(n_1n_2n)$ are much less than $p$,
the operators $b(n_1n_2nk)$ and $b(n_1n_2nk)^*$ should be
interpreted as physical operators describing annihilation and
creation of antiparticles, respectively.

In GFQT the $(b,b^*)$ operators cannot be independent of the
$(a,a^*)$ operators since the latter are defined for all
possible quantum numbers. Therefore the $(b,b^*)$ operators
should be expressed in terms of the $(a,a^*)$ ones. We can
implement the above idea if the operator $b(n_1n_2nk)$ is
defined in such a way that it is proportional to $a({\tilde
n}_1,{\tilde n}_2,n,k)^*$ and hence $b(n_1n_2nk)^*$ is
proportional to $a({\tilde n}_1,{\tilde n}_2,n,k)$.

Since Equation (\ref{25}) should now be considered in $F_p$, it
follows from the well known Wilson theorem $(p-1)!=-1$ in $F_p$
(see e.g., \cite{VDW,Ireland,Davenport}) that
\begin{equation}
F(n_1n_2nk)F({\tilde n}_1{\tilde n}_2nk) = (-1)^s
\label{46}
\end{equation}
We now define the $b$-operators as
\begin{equation}
a(n_1n_2nk)^*=\eta(n_1n_2nk) b({\tilde n}_1{\tilde n}_2nk)/
F({\tilde n}_1{\tilde n}_2nk)
\label{47}
\end{equation}
where $\eta(n_1n_2nk)$ is some function. As a consequence,
\begin{eqnarray}
&a(n_1n_2nk)=\bar{\eta}(n_1n_2nk) b({\tilde n}_1{\tilde n}_2nk)^*/
F({\tilde n}_1{\tilde n}_2nk)\nonumber\\
&b(n_1n_2nk)^*=a({\tilde n}_1{\tilde n}_2nk)
F(n_1n_2nk)/{\bar \eta}({\tilde n}_1{\tilde n}_2nk)\nonumber\\
&b(n_1n_2nk)=a({\tilde n}_1{\tilde n}_2nk)^*
F(n_1n_2nk)/\eta({\tilde n}_1{\tilde n}_2nk)
\label{48}
\end{eqnarray}

Equations (\ref{47}) and (\ref{48}) define a relation between
the sets $(a,a^*)$ and $(b,b^*)$. Although our motivation was
to replace the $(a,a^*)$ operators by the $(b,b^*)$ ones only
for the nonphysical values of the quantum numbers, we can
consider this definition for all the values of $(n_1n_2nk)$.
The transformation described by Equations ({\ref{47}) and
(\ref{48}) can also be treated as a special case of the
Bogolubov transformation discussed in a wide literature on
many-body theory (see e.g., Chapter 10 in Reference
\cite{Walecka} and references therein).

We have not discussed yet what exact definition of the physical
and nonphysical quantum numbers should be. This problem will be
discussed in Section \ref{S5}. However, one might accept

{\it Physical-nonphysical states assumption: Each set of
quantum numbers $(n_1n_2nk)$ is either physical or unphysical.
If it is physical then the set $({\tilde n}_1{\tilde n}_2nk)$
is unphysical and vice versa.}

With this assumption we can conclude from Equations (\ref{47})
and (\ref{48}) that if some operator $a$ is physical then the
corresponding operator $b^*$ is unphysical and vice versa while
if some operator $a^*$ is physical then the corresponding
operator $b$ is unphysical and vice versa.

We have no ground to think that the set of the $(a,a^*)$
operators is more fundamental than the set of the $(b,b^*)$
operators and vice versa. Therefore the question arises whether
the $(b,b^*)$ operators satisfy the relations (\ref{38}) or
(\ref{39}) in the case of anticommutation or commutation
relations, respectively and whether the operators $A_i$ (see
Equation (\ref{44})) have the same form in terms of the
$(a,a^*)$ and $(b,b^*)$ operators. In other words, if the
$(a,a^*)$ operators in Equation (\ref{44}) are expressed in
terms of the $(b,b^*)$ ones then the problem arises whether
\begin{eqnarray}
A_i=\sum A_i(n_1'n_2'n'k',n_1n_2nk)b(n_1'n_2'n'k')^*b(n_1n_2nk)/Norm(n_1n_2nk)
\label{49}
\end{eqnarray}
is valid. It is natural to accept the following

{\it Definition of the AB symmetry: If the $(b,b^*)$ operators
satisfy Equation (\ref{39}) in the case of anticommutators or
Equation (\ref{40}) in the case of commutators and all the
representation operators (\ref{44}) in terms of the $(b,b^*)$
operators have the form (\ref{49}) then it is said that the AB
symmetry is satisfied.}

To prove the AB symmetry we will first investigate whether
Equations (\ref{39}) and (\ref{40}) follow from Equations
(\ref{37}) and (\ref{38}), respectively. As follows from
Equations (\ref{46}-\ref{48}), Equation (\ref{39}) follows from
Equation (\ref{37}) if
\begin{equation}
\eta(n_1n_2nk) {\bar \eta}(n_1,n_2,nk)=(-1)^s
\label{50}
\end{equation}
while Equation (\ref{40}) follows from Equation (\ref{38}) if
\begin{equation}
\eta(n_1n_2nk) {\bar \eta}(n_1,n_2,nk)=(-1)^{s+1}
\label{51}
\end{equation}
We now represent $\eta(n_1n_2nk)$ in the form
\begin{equation}
\eta(n_1n_2nk)=\alpha f(n_1n_2nk)
\label{52}
\end{equation}
where $f(n_1n_2nk)$ should satisfy the condition
\begin{equation}
f(n_1n_2nk) {\bar f}(n_1,n_2,nk)=1
\label{53}
\end{equation}
Then $\alpha$ should be such that
\begin{equation}
\alpha {\bar \alpha}=\pm (-1)^s
\label{54}
\end{equation}
where the plus sign refers to anticommutators and the minus
sign to commutators, respectively. If the normal
spin-statistics connection is valid, \emph{i.e.}, we have
anticommutators for odd values of $s$ and commutators for even
ones then the r.h.s. of Equation (\ref{54}) equals -1 while in
the opposite case it equals 1. In Section \ref{S7}, Equation
(\ref{54}) is discussed in detail and for now we assume that
solutions of this relation exist.

A direct calculation using the explicit expressions
(\ref{30}-\ref{35}) for the matrix elements shows
that if $\eta(n_1n_2nk)$ is given by
Equation (\ref{52}) and
\begin{equation}
f(n_1n_2nk)=(-1)^{n_1+n_2+n}
\label{55}
\end{equation}
then the AB symmetry is valid regardless of whether the normal
spin-statistics connection is valid or not (the details of
calculations can be found in Reference \cite{hep}).

As noted in Section \ref{S1}, elementary particle can be
defined either in the spirit of QFT or in terms of IRs. We now
can give another definition: a particle is elementary if its
operators $(a,a^*)$ (or $(b,b^*)$) are used for describing our
system in the Fock space. A difference between this definition
and that in terms of IRs is clear in the case of massless
particles: they are described by IRs but are treated as
elementary or not depending on whether the description in the
Fock space involves the $(a,a^*)$ operators for the massless
particles or singletons.

\section{Physical and Nonphysical States}
\label{S5}

\begin{sloppypar}
The operator $a(n_1n_2nk)$ can be the physical annihilation
operator only if it annihilates the vacuum vector $\Phi_0$.
Then if the operators $a(n_1n_2nk)$ and $a(n_1n_2nk)^*$ satisfy
the relations (\ref{37}) or (\ref{38}), the vector
$a(n_1n_2nk)^* \Phi_0$ has the meaning of the one-particle
state. The same can be said about the operators $b(n_1n_2nk)$
and $b(n_1n_2nk)^*$. For these reasons in standard theory it is
required that the vacuum vector should satisfy the conditions
(\ref{41}). Then the elements
\begin{equation}
\Phi_+(n_1n_2nk)=a(n_1n_2nk)^*\Phi_0\quad
\Phi_-(n_1n_2nk)=b(n_1n_2nk)^*\Phi_0
\label{56}
\end{equation}
have the meaning of one-particle states for particles and
antiparticles, respectively.
\end{sloppypar}

However, if one requires the condition (\ref{41}) in GFQT, then
it is obvious from Equations (\ref{47}) and Equation (\ref{48})
that the elements defined by Equation (\ref{56}) are null
vectors. Note that in standard approach the AdS energy is
always greater than $m$ while in GFQT the AdS energy is not
positive definite. We can therefore try to modify Equation
(\ref{41}) as follows. Suppose that {\it Physical-nonphysical
states assumption} (see Section \ref{S4}) can be substantiated.
Then we can break the set of elements $(n_1n_2nk)$ into two
nonintersecting parts with the same number of elements, $S_+$
and $S_-$, such that if $(n_1n_2nk)\in S_+$ then $({\tilde
n}_1{\tilde n}_2nk)\in S_-$ and vice versa. Then, instead of
the condition (\ref{41}) we require
\begin{equation}
a(n_1n_2nk)\Phi_0=b(n_1n_2nk)\Phi_0=0\quad
\forall\,\, (n_1,n_2,n,k)\in S_+
\label{57}
\end{equation}
In that case the elements defined by Equation (\ref{56}) will
indeed have the meaning of one-particle states for
$(n_1n_2nk)\in S_+$.

It is clear that if we wish to work with the full set of
elements $(n_1n_2nk)$ then, as follows from Equations
(\ref{47}) and (\ref{48}), the operators $(b,b^*)$ are
redundant and we can work only with the operators $(a,a^*)$.
However, if one works with the both sets, $(a,a^*)$ and
$(b,b^*)$ then such operators can be independent of each other
only for a half of the elements $(n_1n_2nk)$.

\begin{sloppypar}
Regardless of how the sets $S_+$ and $S_-$ are defined, the
{\it Physical-nonphysical states assumption} cannot be
consistent if there exist quantum numbers $(n_1n_2nk)$ such
that $n_1={\tilde n}_1$ and $n_2={\tilde n}_2$. Indeed, in that
case the sets $(n_1n_2nk)$ and $({\tilde n}_1{\tilde n}_2nk)$
are the same what contradicts the assumption that each set
$(n_1n_2nk)$ belongs either to $S_+$ or $S_-$.
\end{sloppypar}

Since the replacements $n_1\rightarrow {\tilde n}_1$ and
$n_2\rightarrow {\tilde n}_2$ change the signs of the
eigenvalues of the $h_1$ and $h_2$ operators (see Section
\ref{S4}), the condition that that $n_1={\tilde n}_1$ and
$n_2={\tilde n}_2$ should be valid simultaneously implies that
the eigenvalues of the operators $h_1$ and $h_2$ should be
equal to zero simultaneously. Recall that (see Section
\ref{S2}) if one considers IR of the sp(2) algebra and treats
the eigenvalues of the diagonal operator $h$ not as elements of
$F_p$ but as integers, then they take the values of
$q_0,q_0+2,...2p-q_0-2,2p-q_0$. Therefore the eigenvalue is
equal to zero in $F_p$ only if it is equal to $p$ when
considered as an integer. Since $m=q_1+q_2$ and the AdS energy
is $E=h_1+h_2$, the above situation can take place only if the
energy considered as an integer is equal to 2p. It now follows
from Equation (\ref{12}) that the energy can be equal to $2p$
only if $m$ is even. Since $s=q_1-q_2$, we conclude that $m$
can be even if and only if $s$ is even. In that case we will
necessarily have quantum numbers $(n_1n_2nk)$ such that the
sets $(n_1n_2nk)$ and $({\tilde n}_1{\tilde n}_2nk)$ are the
same and therefore the {\it Physical-nonphysical states
assumption} is not valid. On the other hand, if $s$ is odd
(\emph{i.e.}, half-integer in the usual units) then there are no
quantum numbers $(n_1n_2nk)$ such that the sets $(n_1n_2nk)$
and $({\tilde n}_1{\tilde n}_2nk)$ are the same.

Our conclusion is as follows: {\it If the separation of states
should be valid for any quantum numbers then the spin $s$
should be necessarily odd.} In other words, if the notion of
particles and antiparticles is absolute then elementary
particles can have only a half-integer spin in the usual units.

In view of the above observations it seems natural to implement
the {\it Physical-nonphysical states assumption} as follows.
{\it If the quantum numbers $(n_1n_2nk)$ are such that
$m+2(n_1+n_2+n) < 2p$ then the corresponding state is physical
and belongs to $S_+$, otherwise the state is unphysical and
belongs to $S_-$.} However, one cannot guarantee that there are
no other reasonable implementations.

\section{AdS Symmetry Breaking}
\label{breaking}

In view of the above discussion, our next goal is the
following. We should take the operators in the form (\ref{44})
and replace the $(a,a^*)$ operators by the $(b,b^*)$ ones only
if $(n_1n_2nk)\in S_-$. Then a question arises whether we will
obtain the standard result (\ref{42}) where a sum is taken only
over values of $(n_1n_2nk)\in S_+$. The fact that we have
proved the AB symmetry does not guarantee that this is the case
since the AB symmetry implies that the replacement has been
made for all the quantum numbers, not only half of them.
However, the derivation of the AB symmetry shows that for the
contribution of such quantum numbers that $(n_1n_2nk)\in S_+$
and $(n_1'n_2'n'k')\in S_+$ we will indeed have the result
(\ref{42}) up to some constants. This derivation also
guarantees that if we consider the action of the operators on
states described by physical quantum numbers and the result of
the action also is a state described by physical quantum
numbers then on such states the correct commutation relations
are satisfied. A problem arises whether they will be satisfied
for transitions between physical and nonphysical quantum
numbers.

Let $A(a_1^{'})$ be the secondly quantized operator
corresponding to $a_1^{'}$ and $A(a_1^{"})$ be the secondly
quantized operator corresponding to $a_1^{"}$. Consider the
action of these operators on the state
$\Phi=a(n_1n_2nk)^*\Phi_0$ such that $(n_1n_2nk)\in S_+$ but
$(n_1+1,n_2nk)\in S_-$. As follows from Equations (\ref{9}) and
(\ref{30}), we should have
\begin{equation}
[A(a_1^{'}),A(a_1^{"})]\Phi =[Q_1(n,k)+2n_1]\Phi
\label{correct}
\end{equation}
As follows from Equations (\ref{31}) and (\ref{47}),
$A(a_1^{"})\Phi=a(n_1+1,n_2nk)^*\Phi_0$. Since
$(n_1+1,n_2nk)\in S_-$, we should replace $a(n_1+1,n_2nk)^*$ by
an operator proportional to $b({\tilde n}_1-1,{\tilde n}_2nk)$
and then, as follows from Equation (\ref{41}),
$A(a_1^{"})\Phi=0$. Now, by using Equations (\ref{31}) and
(\ref{47}), we get
\begin{equation}
[A(a_1^{'}),A(a_1^{"})]\Phi =n_1[Q_1(n,k)+n_1-1]\Phi
\label{incorrect}
\end{equation}
Equations (\ref{correct}) and (\ref{incorrect}) are
incompatible with each other and we conclude that our procedure
breaks the AdS symmetry for transitions between physical and
nonphysical states.

We conclude that if, by analogy with standard theory, one
wishes to interpret modular IRs of the dS algebra in terms of
particles and antiparticles then the commutation relations of
the dS algebra will be broken. This does not mean that such a
possibility contradicts the existing knowledge since they will
be broken only at extremely high dS energies of order $p$. At
the same time, a possible point of view is that since we
started from the symmetry algebra, we should not sacrifice
symmetry because we don't know other ways of interpreting IRs.
The mathematical structure of IRs indicates that they describe
objects characterized by quantum numbers $(n_1n_2nk)$ and
breaking this set of quantum numbers into $S_+$ and $S_-$ is
only an approximation valid at not very high energies. If we
accept this point of view then there is no need to require that
if quantum numbers $(n_1,n_2nk)$ are physical then the numbers
$({\tilde n}_1{\tilde n}_2nk)$ are unphysical and vice versa.
For example, we can exclude such quantum numbers that
$n_1={\tilde n}_1$ and $n_2={\tilde n}_2$ and therefore a
description in terms of particles and antiparticles will be
valid in the case of even $s$ too.

\section{Dirac Vacuum Energy Problem}
\label{S6}

The Dirac vacuum energy problem is discussed in practically
every textbook on QFT. In its simplified form it can be
described as follows. Suppose that the energy spectrum is
discrete and $n$ is the quantum number enumerating the states.
Let $E(n)$ be the energy in the state $n$. Consider the
electron-positron field. As a result of quantization one gets
for the energy operator
\begin{equation}
E = \sum_n E(n)[a(n)^*a(n)-b(n)b(n)^*]
\label{58}
\end{equation}
where $a(n)$ is the operator of electron annihilation in the
state $n$, $a(n)^*$ is the operator of electron creation in the
state $n$, $b(n)$ is the operator of positron annihilation in
the state $n$ and $b(n)^*$ is the operator of positron creation
in the state $n$. It follows from this expression that only
anticommutation relations are possible since otherwise the
energy of positrons will be negative. However, if
anticommutation relations are assumed, it follows from Equation
(\ref{58}) that
\begin{equation}
E = \{\sum_n E(n)[a(n)^*a(n)+b(n)^*b(n)]\}+E_0
\label{59}
\end{equation}
where $E_0$ is some infinite negative constant. Its presence
was a motivation for developing Dirac's hole theory. In the
modern approach it is usually required that the vacuum energy
should be zero. This can be obtained by assuming that all
operators should be written in the normal form. However, this
requirement is not quite consistent since the result of
quantization is Equation (\ref{58}) where the positron
operators are not written in that form (see also the discussion
in Section \ref{S4}).

Consider now the AdS energy operator $M^{05}=h_1+h_2$ in GFQT.
As follows from Equations (\ref{30}) and (\ref{45})
\begin{eqnarray}
M^{05}=\sum [m+2(n_1+n_2+n)]a(n_1n_2nk)^*a(n_1n_2nk)/Norm(n_1n_2nk)
\label{60}
\end{eqnarray}
where the sum is taken over all possible quantum numbers
$(n_1n_2nk)$. As noted in the preceding section, one could try
to interpret this operator in terms of particles and
antiparticles by replacing only the nonphysical $(a,a^*)$
operators by the physical $(b,b^*)$ ones. Then $M^{05}$ will be
represented in terms of physical operators only. As noted in
the preceding section, it is not clear whether such a procedure
is physical or not. Nevertheless it is interesting to see
whether the vacuum energy can be calculated in GFQT and whether
this will shed light on the problem of infinities in standard
QFT.

As follows from Equations (\ref{46}-\ref{48}) and
(\ref{52}-\ref{54})
\begin{eqnarray}
&M^{05}=\{\sum_{S_+} [m+2(n_1+n_2+n)]
[a(n_1n_2nk)^*a(n_1n_2nk)+\nonumber\\
&b(n_1n_2nk)^*b(n_1n_2nk)]/Norm(n_1n_2nk)\}+E_{vac}
\label{61}
\end{eqnarray}
where the vacuum energy is given by
\begin{equation}
E_{vac}=\mp \sum_{S_+}  [m+2(n_1+n_2+n)]
\label{62}
\end{equation}
in the cases when the $(b,b^*)$ operators anticommute and
commute, respectively. Note that in contrast with standard
theory, we have represented the result for $M^{05}$ in the
normal form for both, commutation and anticommutation
relations. For definiteness, we will perform calculations for
the case when the operators anticommute and the value of $s$ is
odd.

Consider first the sum in Equation (\ref{62}) when the values
of $n$ and $k$ are fixed. It is convenient to distinguish the
cases $s > 2k$ and $s<2k$. If $s > 2k$ then, as follows from
Equation (\ref{20}), the maximum value of $n_1$ is such that
$m+2(n+n_1)$ is always less than $2p$. For this reason all the
values of $n_1$ contribute to the sum, which can be written as
\begin{eqnarray}
S_1(n,k) =-\sum_{n_1=0}^{p-q_1-n+k}[(m+2n+2n_1)+(m+2n+2n_1+2)+...+(2p-1)]
\label{63}
\end{eqnarray}
A simple calculation shows that the result can be represented as
\begin{equation}
S_1(n,k)=\sum_{n_1=1}^{p-1}n_1^2-\sum_{n_1=1}^{n+(m-3)/2}n_1^2-
\sum_{n_1=1}^{(s-1)/2-k}n_1^2
\label{65}
\end{equation}
where the last sum should be taken into account only if $(s-1)/2-k\geq 1$.

The first sum in this expression equals $(p-1)p(2p-1)/6$ and,
since we assume that $p\neq 2$ and $p\neq 3$, this quantity is
zero in $F_p$. As a result, $S_1(n,k)$ is represented as a sum
of two terms such that the first one depends only on $n$ and
the second --- only on $k$. Note also that the second term is
absent if $s=1$, \emph{i.e.}, for particles with the spin 1/2 in
the usual units.

Analogously, if $s < 2k$ the result is
\begin{equation}
S_2(n,k)=-\sum_{n_2=1}^{n+(m-3)/2}n_2^2-\sum_{n_2=1}^{k-(s+1)/2}n_2^2
\label{67}
\end{equation}
where the second term should be taken into account only if
$k-(s+1)/2\geq 1$.

We now should calculate the sum
\begin{equation}
S(n)=\sum_{k=0}^{(s-1)/2}S_1(n,k) +\sum_{k=(s+1)/2}^s S_2(n,k)
\label{68}
\end{equation}
and the result is
\begin{eqnarray}
&S(n)=-(s+1)(n+\frac{m-1}{2})[2(n+\frac{m-1}{2})^2-\nonumber\\
&3(n+\frac{m-1}{2})+1]/6-(s-1)(s+1)^2(s+3)/96
\label{69}
\end{eqnarray}
Since the value of $n$ is in the range $[0,n_{max}]$, the final
result is
\begin{equation}
E_{vac}=\sum_{n=0}^{n_{max}}S(n)=(m-3)(s-1)(s+1)^2(s+3)/96
\label{70}
\end{equation}
since in the massive case $n_{max}=p+2-m$.

Our final conclusion in this section is that {\it if $s$ is odd
and the separation of states into physical and nonphysical ones
is accomplished as in Section \ref{S5} then $E_{vac}=0$ only if
$s=1$ (\emph{i.e.}, $s=1/2$ in the usual units)}. This result
shows that since the rules of arithmetic in Galois fields are
different from that for real numbers, it is possible that
quantities which are infinite in standard theory will be zero
in GFQT.

\section{Neutral Particles and Spin-Statistics Theorem}
\label{S7}

In this section we will discuss the relation between the
$(a,a^*)$ and $(b,b^*)$ operators only for all quantum numbers
(\emph{i.e.}, in the spirit of the AB-symmetry) and therefore
the results are valid regardless of whether the separation of
states into $S_+$ and $S_-$ can be justified or not (see the
discussion in Section \ref{breaking}).

The nonexistence of neutral elementary particles in GFQT is one
of the most striking differences between GFQT and standard
theory. One could give the following definition of neutral
particle:
\begin{itemize}
\item i) it is a particle coinciding with its
antiparticle
\item ii) it is a particle which does not coincide
with its antiparticle but they have the same properties
\end{itemize}
In standard theory only i) is meaningful since neutral
particles are described by real (not complex) fields and this
condition is required by Hermiticity. One might think that the
definition ii) is only academic since if a particle and its
antiparticle have the same properties then they are
indistinguishable and can be treated as the same. However, the
cases i) and ii) are essentially different from the operator
point of view. In the case i) only the $(a,a^*)$ operators are
sufficient for  describing the operators (\ref{42}) in standard
theory. This is the reflection of the fact that the real field
has the number of degrees of freedom twice as less as the
complex field. On the other hand, in the case ii) both
$(a,a^*)$ and $(b,b^*)$ operators are required, \emph{i.e.}, in
standard theory such a situation is described by a complex
field. Nevertheless, the case ii) seems to be rather odd: it
implies that there exists a quantum number distinguishing a
particle from its antiparticle but this number is not
manifested experimentally. We now consider whether the
conditions i) or ii) can be implemented in GFQT.

Since each operator $a$ is proportional to some operator $b^*$
and vice versa (see Equations (\ref{47}) and (\ref{48})), it is
clear that if the particles described by the operators
$(a,a^*)$ have a nonzero charge then the particles described by
the operators $(b,b^*)$ have the opposite charge and the number
of operators cannot be reduced. However, if all possible
charges are zero, one could try to implement i) by requiring
that each $b(n_1n_2nk)$ should be proportional to $a(n_1n_2nk)$
and then $a(n_1n_2nk)$ will be proportional to $a({\tilde
n}_1,{\tilde n}_2,nk)^*$. In this case the operators $(b,b^*)$
will not be needed at all.

Suppose, for example, that the operators $(a,a^*)$ satisfy the
commutation relations (\ref{38}). In that case the operators
$a(n_1n_2nk)$ and $a(n_1'n_2'n'k')$ should commute if the sets
$(n_1n_2nk)$ and $(n_1'n_2'n'k')$ are not the same. In
particular, one should have $[a(n_1n_2nk), a({\tilde
n}_1{\tilde n}_2nk)]=0$ if either $n_1\neq {\tilde n}_1$ or
$n_2\neq {\tilde n}_2$. On the other hand, if $a({\tilde
n}_1{\tilde n}_2nk)$ is proportional to $a(n_1n_2nk)^*$, it
follows from Equation (\ref{38}) that the commutator cannot be
zero. Analogously one can consider the case of anticommutators.

The fact that the number of operators cannot be reduced is also
clear from the observation that the $(a,a^*)$ or $(b,b^*)$
operators describe an irreducible representation in which the
number of states (by definition) cannot be reduced. Our
conclusion is that in GFQT the definition of neutral particle
according to i) is fully unacceptable.

Consider now whether it is possible to implement the definition
ii) in GFQT. Recall that we started from the operators
$(a,a^*)$ and defined the operators $(b,b^*)$ by means of
Equation (\ref{47}). Then the latter satisfy the same
commutation or anticommutation relations as the former and the
AB symmetry is valid. Does it mean that the particles described
by the operators $(b,b^*)$ are the same as the ones described
by the operators $(a,a^*)$? If one starts from the operators
$(b,b^*)$ then, by analogy with Equation (\ref{47}), the
operators $(a,a^*)$ can be defined as
\begin{equation}
b(n_1n_2nk)^*=\eta'(n_1n_2nk) a({\tilde n}_1{\tilde n}_2nk)/
F({\tilde n}_1{\tilde n}_2nk)
\label{72}
\end{equation}
where $\eta'(n_1n_2nk)$ is some function. By analogy with the
consideration in Section \ref{S4} one can show that
\begin{equation}
\eta'(n_1n_2nk)=\beta (-1)^{n_1+n_2+n}\quad
\beta {\bar \beta}=\mp 1
\label{73}
\end{equation}
where the minus sign refers to the normal spin-statistics
connection and the plus to the broken one.

As follows from Equations (\ref{47}), (\ref{50}-\ref{53}),
(\ref{72}), (\ref{73}) and the definition of the quantities
${\tilde n}_1$ and ${\tilde n}_2$ in Section \ref{S4}, the
relation between the quantities $\alpha$ and $\beta$ is $\alpha
{\bar \beta}=1$. Therefore, as follows from Equation
(\ref{73}), there exist only two possibilities, $\beta = \mp
\alpha$, depending on whether the normal spin-statistics
connection is valid or not. We conclude that the broken
spin-statistics connection implies that $\alpha{\bar
\alpha}=\beta{\bar\beta}=1$ and $\beta=\alpha$ while the normal
spin-statistics connection implies that $\alpha{\bar
\alpha}=\beta{\bar\beta}=-1$ and $\beta=-\alpha$. Since in the
first case there exist solutions such that $\alpha=\beta$ (e.g.,
$\alpha = \beta = 1$), the particle and its antiparticle can be
treated as neutral in the sense of the definition ii). Since
such a situation is clearly unphysical, one might treat the
spin-statistics theorem as a requirement excluding neutral
particles in the sense ii).

\begin{sloppypar}
We now consider another possible treatment of the
spin-statistics theorem, which seems to be much more
interesting. In the case of normal spin-statistics connection
we have that
\begin{equation}
\alpha {\bar \alpha}=-1
\label{77}
\end{equation}
and the problem arises whether solutions of this relation
exist. Such a relation is obviously impossible in standard
theory.
\end{sloppypar}

As noted in Section \ref{S1}, $-1$ is a quadratic residue in
$F_p$ if $p=1\,\, (mod\,\, 4)$ and a quadratic non-residue in
$F_p$ if $p=3\,\, (mod\,\, 4)$. For example, $-1$ is a
quadratic residue in $F_5$ since $2^2=-1\,\, (mod\,\, 5)$ but
in $F_7$ there is no element $a$ such that $a^2=-1\,\, (mod\,\,
7)$. We conclude that if $p=1\,\, (mod\,\, 4)$ then Equation
(\ref{77}) has solutions in $F_p$ and in that case the theory
can be constructed without any extension of $F_p$.

Consider now the case $p=3\,\, (mod\,\, 4)$. Then Equation
(\ref{77}) has no solutions in $F_p$ and it is necessary to
consider this equation in an extension of $F_p$ (\emph{i.e.},
there is no ``real'' version of GFQT). The minimum extension is
obviously $F_{p^2}$ and therefore the problem arises whether
Equation (\ref{77}) has solutions in $F_{p^2}$.

It is well known \cite{VDW,Ireland,Davenport} that any Galois field without its
zero element is a cyclic multiplicative group. Let $r$ be a
primitive root, \emph{i.e.}, the element such that any nonzero
element of $F_{p^2}$ can be represented as $r^k$
$(k=1,2,...,p^2-1)$. It is also well known that the only
nontrivial automorphism of $F_{p^2}$ is $\alpha\rightarrow
{\bar \alpha}=\alpha^p$. Therefore if $\alpha =r^k$ then
$\alpha{\bar \alpha}= r^{(p+1)k}$. On the other hand, since
$r^{(p^2-1)}=1$, $r^{(p^2-1)/2}=-1$. Therefore there exists at
least a solution with $k=(p-1)/2$.

Our conclusion is that {\it if $p=3\,\, (mod\,\, 4 )$ then the
spin-statistics theorem implies that the field $F_p$ should
necessarily be  extended and the minimum possible extension is
$F_{p^2}$}. Therefore the spin-statistics theorem can be
treated as a requirement that GFQT should be based on $F_{p^2}$
and standard theory should be based on complex numbers.

Let us now discuss a different approach to the AB symmetry. A
desire to have operators which can be interpreted as those
relating separately to particles and antiparticles is natural
in view of our experience in standard approach. However, one
might think that in the spirit of GFQT there is no need to have
separate operators for particles and antiparticles since they
are different states of the same object. We can therefore
reformulate the AB symmetry in terms of only $(a,a^*)$
operators as follows. Instead of Equations (\ref{47}) and
(\ref{48}), we consider a {\it transformation} defined as
\begin{eqnarray}
&a(n_1n_2nk)^*\rightarrow \eta(n_1n_2nk)
a({\tilde n}_1{\tilde n}_2nk)/
F({\tilde n}_1{\tilde n}_2nk)\nonumber\\
&a(n_1n_2nk)\rightarrow \bar{\eta}(n_1n_2nk)
a({\tilde n}_1{\tilde n}_2nk)^*/
F({\tilde n}_1{\tilde n}_2nk)
\label{78}
\end{eqnarray}
Then the AB symmetry can be formulated as a requirement that
physical results should be invariant under this transformation.

\begin{sloppypar}
Let us now apply the AB transformation twice. Then, by analogy
with the derivation of Equation (\ref{54}), we get
\begin{equation}
a(n_1n_2nk)^*\rightarrow \mp a(n_1n_2nk)^*\quad
a(n_1n_2nk)\rightarrow \mp a(n_1n_2nk)
\label{79}
\end{equation}
for the normal and broken spin-statistic connections,
respectively. Therefore, as a consequence of the
spin-statistics theorem, any particle (with the integer or
half-integer spin) has the AB$^2$ parity equal to $-1$.
Therefore in GFQT any interaction can involve only an even
number of creation and annihilation operators. In particular,
this is additional demonstration of the fact that in GFQT the
existence of neutral elementary particles is incompatible with
the spin-statistics theorem.
\end{sloppypar}

\section{Modular IRs of the osp(1,4) Superalgebra}
\label{SS}

If one accepts supersymmetry then the results on modular IRs of
the so(2,3) algebra can be generalized by considering modular
IRs of the osp(1,4) superalgebra. Representations of the
osp(1,4) superalgebra have several interesting distinctions
from representations of the Poincare superalgebra. For this
reason we first briefly mention some well known facts about the
latter representations (see e.g Reference \cite{Wein-super} for
details).

Representations of the Poincare superalgebra are described by
14 operators. Ten of them are the well known representation
operators of the Poincare algebra---four momentum operators and
six representation operators of the Lorentz algebra, which
satisfy the well known commutation relations. In addition,
there also exist four fermionic operators. The anticommutators
of the fermionic operators are linear combinations of the
momentum operators, and the commutators of the fermionic
operators with the Lorentz algebra operators are linear
combinations of the fermionic operators. In addition, the
fermionic operators commute with the momentum operators.

From the formal point of view, representations of the osp(1,4)
superalgebra are also described by 14 operators --- ten
representation operators of the so(2,3) algebra and four
fermionic operators. There are three types of relations: the
operators of the so(2,3) algebra commute with each other as
usual (see Section \ref{S3}), anticommutators of the fermionic
operators are linear combinations of the so(2,3) operators and
commutators of the latter with the fermionic operators are
their linear combinations. However, in fact representations of
the osp(1,4) superalgebra can be described exclusively in terms
of the fermionic operators. The matter is as follows. In the
general case the anticommutators of four operators form ten
independent linear combinations. Therefore, ten bosonic
operators can be expressed in terms of fermionic ones. This is
not the case for the Poincare superalgebra since the Poincare
algebra operators are obtained from the so(2,3) ones by
contraction. One can say that the representations of the
osp(1,4) superalgebra is an implementation of the idea that
supersymmetry is the extraction of the square root from the
usual symmetry (by analogy with the well known treatment of the
Dirac equation as a square root from the Klein-Gordon one).

\begin{sloppypar}
We denote the fermionic operators of the osp(1,4) superalgebra
as $(d_1,d_2,d_1^*,d_2^*)$ where the $^*$ means the Hermitian
conjugation as usual. They should satisfy the following
relations. If $(A,B,C)$ are any fermionic operators, [...,...]
is used to denote a commutator and $\{...,...\}$ to denote an
anticommutator then
\begin{equation}
[A,\{ B,C\} ]=F(A,B)C + F(A,C)B
\label{S30}
\end{equation}
where the form $F(A,B)$ is skew symmetric, $F(d_j,d_j^*)=1$
$(j=1,2)$ and the other independent values of $F(A,B)$ are
equal to zero. The fact that the representation of the osp(1,4)
superalgebra is fully defined by Equation (\ref{S30}) and the
properties of the form $F(.,.)$, shows that osp(1,4) is a
special case of the superalgebra.
\end{sloppypar}

We can now {\bf define} the so(2,3) generators as follows:
\begin{eqnarray}
&b'=\{d_1,d_2\}\quad b"=\{d_1^*,d_2^*\}\quad
L_+=\{d_2,d_1^*\}\quad L_-=\{d_1,d_2^*\}\nonumber\\
&a_j'=(d_j)^2\quad a_j"=(d_j^*)^2\quad
h_j=\{d_j,d_j^*\} \quad (j=1,2)
\label{S31}
\end{eqnarray}
Then by using Equation (\ref{S30}) and the properties of the
form $F(.,.)$, one can show by direct calculations that so
defined operators satisfy the commutation relations
(\ref{9}-\ref{11}). This result can be treated as a fact that
the operators of the so(2,3) algebra are not fundamental, only
the fermionic operators are.

By analogy with the construction of IRs of the osp(1,4)
superalgebra in standard theory \cite{Heidenreich}, we require
the existence of the cyclic vector $e_0$ satisfying the
conditions (compare with Equation (\ref{15})):
\begin{eqnarray}
d_je_0=L_+e_0=0 \quad h_je_0=q_je_0\quad (e_0,e_0)\neq 0\quad (j=1,2)
\label{S32}
\end{eqnarray}
The full representation space can be obtained by successively
acting by the fermionic operators on $e_0$ and taking all
possible linear combinations of such vectors.

\begin{sloppypar}
We use $E$ to denote an arbitrary linear combination of the
vectors $(e_0,d_1^*e_0,d_2^*e_0,d_2^*d_1^*e_0)$. Our next goal
is to prove a statement analogous to that in Reference \cite{Heidenreich}:
\end{sloppypar}

{\it Statement 1}: Any vector from the representation
space can be represented as a linear combination of the
elements $O_1O_2...O_nE$ where $n=0,1,...$ and $O_i$ is an operator
of the so(2,3) algebra.

The first step is to prove a simple

{\it Lemma:} If $D$ is any fermionic operator then DE is a
linear combination of elements $E$ and $OE$ where $O$ is an operator
of the so(2,3) algebra.

The proof is by a straightforward check using Equations
(\ref{S30}-\ref{S32}). For example,
$$d_1^*(d_2^*d_1^*e_0)=\{d_1^*,d_2^*\}d_1^*e_0-d_2^*a_1"e_0=
b"d_1^*e_0-a_1"d_2^*e_0\,\, $$

To prove Statement 1 we define the height of a linear
combination of the elements $O_1O_2...O_nE$ as the maximum sum
of powers of the fermionic operator in this element. For
example, since every operator of the so(2,3) algebra is
composed of two fermionic operator, the height of the element
$O_1O_2...O_nE$ equals $2n+2$ if $E$ contains $d_2^*d_1^*e_0$,
equals $2n+1$ if $E$ does not contain $d_2^*d_1^*e_0$ but
contains either $d_1^*e_0$ or $d_2^*e_0$ and equals $2n$ if $E$
contains only $e_0$.

We can now prove Statement 1 by induction. The elements with
the heights 0,1 and 2 obviously have the required form since,
as follows from Equation (\ref{S31}),
$d_1^*d_2^*e_0=b"e_0-d_2^*d_1^*e_0$. Let us assume that
Statement 1 is correct for all elements with the heights $\leq
N$. Every element with the height $N+1$ can be represented as
$Dx$ where $x$ is an element with the height $N$. If
$x=O_1O_2...O_nE$ then by using Equation (\ref{S30}) we can
represent $Dx$ as $Dx=O_1O_2...O_nDE+y$ where the height of the
element $y$ is equal to $N-1$. As follows from the induction
assumption, $y$ has the required form, and, as follows from
Lemma, $DE$ is a linear combination of the elements $E$ and
$OE$. Therefore Statement 1 is proved.

As follows from Equations (\ref{S30}) and (\ref{S31}),
\begin{eqnarray}
[d_j,h_j]=d_j\quad [d_j^*,h_j]=-d_j^*\quad [d_j,h_l]=[d_j^*,h_l]=0\quad (j,l=1,2\,\, j\neq l)
\label{S33}
\end{eqnarray}
It follows from these expressions that if $x$ is such that
$h_jx=\alpha_jx$ $(j=1,2)$ then $d_1^*x$ is the eigenvector of
the operators $h_j$ with the eigenvalues
$(\alpha_1+1,\alpha_2)$, $d_2^*x$ - with the eigenvalues
$(\alpha_1,\alpha_2+1)$, $d_1x$ - with the eigenvalues
$(\alpha_1-1,\alpha_2)$, and $d_2x$ - with the eigenvalues
$\alpha_1,\alpha_2-1$.

Let us assume that $q_2\geq 1$ and $q_1\geq q_2$. We again use
$m$ to denote $q_1+q_2$ and $s$ to denote $q_1-q_2$. Statement
1 obviously remains valid if we now assume that $E$ contains
linear combinations of $(e_0,e_1,e_2,e_3)$ where
\begin{eqnarray}
e_1=d_1^*e_0\quad e_2=d_2^*e_0-\frac{1}{s+1}L_-e_1\quad e_3=(d_2^*d_1^*e_0-\frac{q_1-1}{m-2}b"+
\frac{1}{m-2}a_1"L_-)e_0
\label{S34}
\end{eqnarray}
We assume for simplicity that $(e_0,e_0)=1$. Then
it can be shown by direct calculations using Equations
(\ref{S30}-\ref{S32}) that
\begin{equation}
(e_1,e_1)=q_1 \quad (e_2,e_2)=\frac{s(q_2-1)}{s+1}
\quad (e_3,e_3)=\frac{q_1(q_2-1)(m-1)}{m-2}
\label{S35}
\end{equation}

As follows from Equations (\ref{S30}-\ref{S33}), $e_0$
satisfies Equation (\ref{15}) and $e_1$ satisfies the same
condition with $q_1$ replaced by $q_1+1$. We see that the
representation of the osp(1,4) superalgebra defined by Equation
(\ref{S32}) necessarily contains at least two IRs of the
so(2,3) algebra characterized by the values of the mass and
spin $(m,s)$ and $(m+1,s+1)$, and the cyclic vectors $e_0$ and
$e_1$, respectively.

As follows from Equations (\ref{S30}-\ref{S33}), the vectors $e_2$ and $e_3$ satisfy the conditions
\begin{eqnarray}
&h_1e_2=q_1e_2\quad h_2e_2=(q_2+1)e_2 \quad
h_1e_3=(q_1+1)e_3 \nonumber\\
&h_2e_3=(q_2+1)e_3\quad a_1'e_j=a_2'e_j=b'e_j=L_+e_j=0
\label{S36}
\end{eqnarray}
$(j=2,3)$ and therefore (see Equation (\ref{15})) they are
candidates for being cyclic vectors of IRs of the so(2,3)
algebra if their norm is not equal to zero. As follows from
Equation (\ref{S35}), $(e_2,e_2)\neq 0$ if $s\neq 0$ and
$q_2\neq 1$. Therefore, if these conditions are satisfied,
$e_2$ is the cyclic vector of IR of the so(2,3) algebra
characterized by the values of the mass and spin $(m+1,s-1)$.
Analogously, if $q_2\neq 1$ then $e_3$ is the cyclic vector of
IR of the so(2,3) algebra characterized by the values of the
mass and spin $(m+2,s)$.

As already mentioned, our considerations are similar to those
in Reference \cite{Heidenreich}. Therefore modular IRs of the
osp(1,4) superalgebra can be characterized in the same way as conventional
IRs \cite{Heidenreich,F}:
\begin{itemize}
\item If $q_2>1$ and $s\neq 0$ (massive IRs), the osp(1,4)
supermultiplets contain four IRs of the so(2,3) algebra
characterized by the values of the mass and spin
$(m,s),(m+1,s+1),(m+1,s-1),(m+2,s).$
\item If $q_2>1$ and $s=0$ (collapsed massive IRs), the osp(1,4)
supermultiplets contain three IRs of the so(2,3) algebra
characterized by the values of the mass and spin
$(m,s),(m+1,s+1),(m+2,s).$
\item If $q_2=1$ (massless IRs) the osp(1,4)
supermultiplets contains two IRs of the so(2,3) algebra
characterized by the values of the mass and spin
$(2+s,s),(3+s,s+1)$
\item Dirac supermultiplet containing two Dirac
singletons (see Section \ref{Singletons}).
\end{itemize}

The first three cases have well known analogs of IRs of the
super-Poincare algebra (see e.g., Reference \cite{Wein-super})
while there is no super-Poincare analog of the Dirac
supermultiplet.

Since the space of IR of the superalgebra osp(1,4) is a direct
sum of spaces of IRs of the so(2,3) algebra, for modular IRs of
the osp(1,4) superalgebra one can prove results analogous to
those discussed in the preceding sections. In particular, one
modular IR of the osp(1,4) algebra is a modular analog of both
standard IRs of the osp(1,4) superalgebra with positive and
negative energies. This implies that one modular IR of the
osp(1,4) superalgebra contains both, a superparticle and its
anti-superparticle. It is possible to prove a superanalog of
the AB symmetry and show that the AB symmetries of particles in
the supermultiplet should satisfy certain relations which
impose a restriction on the form of interaction in
supersymmetric theory (see Reference \cite{levsusy}). It is
also possible to show that a separation of states into
superparticles and anti-superparticles encounters the same
problems as in the so(2,3) case. The details of calculations
can be found in Reference \cite{levsusy}.

\section{Discussion}

In the present paper we discuss a quantum theory based on a
Galois field (GFQT). As noted in Section \ref{S1}, GFQT does
not contain infinities at all and all operators are
automatically well defined. In my discussions with physicists,
some of them commented this fact as follows. This is an
approach where a cutoff (the characteristic $p$ of the Galois
field) is introduced from the beginning and for this reason
there is nothing strange in the fact that the theory does not
have infinities. It has a large number $p$ instead and this
number can be practically treated as infinite.

However, the difference between Galois fields and usual numbers
is not only that the former are finite and the latter are
infinite. If the set of usual numbers is visualized as a
straight line from $-\infty$ to $+\infty$ then the simplest
Galois field can be visualized not as a segment of this line
but as a circumference (see Fig. 1 in Section \ref{S1}). This
reflects the fact that in Galois fields the rules of arithmetic
are different and, as a result, GFQT has many unusual features
which have no analogs in standard theory.

The Dirac vacuum energy problem discussed in Section \ref{S6}
is a good illustration of this point. Indeed, in standard
theory the vacuum energy is infinite and, if GFQT is treated
simply as a theory with a cutoff $p$, one would expect the
vacuum energy to be of order $p$. However, since the rules of
arithmetic in Galois fields are different from standard ones,
the result of exact (\emph{i.e.}, non-perturbative) calculation
of the vacuum energy is precisely zero.

The original motivation for investigating GFQT was as follows.
Let us take standard QED in dS or AdS space, write the
Hamiltonian and other operators in angular momentum basis and
replace standard irreducible representations (IRs) for the
electron, positron and photon by corresponding modular IRs. One
might treat this motivation as an attempt to substantiate
standard momentum regularizations (e.g., the Pauli-Villars
regularization) at momenta $p/R$ (where $R$ is the radius of
the Universe). In other terms this might be treated as
introducing fundamental length of order $R/p$. We now discuss
reasons explaining why this naive attempt fails.

The main result of the present paper is that {\it in GFQT the
existence of antiparticles follows from the fact that any
Galois field is finite. Moreover, the very existence of
antiparticles might be an indication that nature is described
rather by a finite field or ring than by complex numbers.} We
believe that this result is not only very important but also
extremely simple and beautiful. A mathematical consideration of
modular IRs is given in Sections \ref{S2}-\ref{Singletons}
while a simple explanation of the above result is as follows.

In standard theory a particle is described by a positive energy
IR where the energy has the spectrum in the range
$[mass,\infty)$. At the same time, the corresponding
antiparticle is associated with a negative energy IR where the
energy has the spectrum in the range $(-\infty,-mass]$.
Consider now the construction of modular IR for some particle.
We again start from the rest state (where energy=mass) and
gradually construct states with higher and higher energies.
However, in such a way we are moving not along a straight line
but along the circumference in Figure 1. Then sooner or later
we will arrive at the point where energy=-mass.

In QFT the fact that a particle and its antiparticle have the
same masses and spins but opposite charges follows from the CPT
theorem, which is a consequence of locality. A question arises
what happens if locality is only an approximation: in that case
the equality of masses, spins \emph{etc.} is exact or
approximate? Consider a simple model when electromagnetic and
weak interactions are absent. Then the fact that the proton and
the neutron have the same masses and spins has nothing to do
with locality; it is only a consequence of the fact that the
proton and the neutron belong to the same isotopic multiplet.
In other words, they are simply different states of the same
object - the nucleon. We see, that in GFQT the situation is
analogous. The fact that a particle and its antiparticle have
the same masses and spins but opposite charges has nothing to
do with locality or non-locality and is simply a consequence of
the fact that they are different states of the same object
since they belong to the same IR.

In standard theory, a particle and its antiparticle are
combined together by a local covariant equation (e.g., the Dirac
equation). We see that in GFQT the idea of the Dirac equation
is implemented without assuming locality but already at the
level of IRs. This automatically explains the existence of
antiparticles, shows that a particle cannot exist by itself
without its antiparticle and that a particle and its
antiparticle are necessarily {\it different} states of the same
object. In particular, there are no elementary particles which
in standard theory are called neutral.

If a particle is characterized by some additive quantum numbers
(e.g., the electric, baryon or lepton charges) then, as follows
from our construction (see Equations (\ref{47}) and
(\ref{48})), the corresponding antiparticle is characterized by
the same quantum numbers but with the opposite sign. In
standard theory such quantum numbers are conserved because IRs
describing a particle and its antiparticle are fully
independent. However, since in GFQT a particle and its
antiparticle belong to the same IR, the problem arises whether
these quantum numbers are exactly conserved. Let us discuss
this problem in greater details.

In quantum theory there is a superselection rule (SSR)
prohibiting states which are superpositions of states with
different electric, baryon or lepton charges. In general, if
states $\psi_1$ and $\psi_2$ are such that there are no
physical operators $A$ such that $(\psi_2,A\psi_1)\neq 0$ then
the SSR says that the state $\psi=\psi_1+\psi_2$ is prohibited.
The meaning of the SSR is now widely discussed (see e.g.,
Reference \cite{Giulini} and references therein). Since the SSR
implies that the superposition principle, which is a key
principle of quantum theory, is not universal, several authors
argue that the SSR should not be present in quantum theory.
Other authors argue that the SSR is only a dynamical principle
since, as a result of decoherence, the state $\psi$ will
quickly disappear and so it cannot be observable.

In our construction, one IR describes an object characterized
by quantum numbers $(n_1n_2nk)$ (see Section \ref{S3}). We have
discussed an interpretation that a half of the numbers are
related to a particle and the another half to the corresponding
antiparticle. However, since those numbers describe the same
IR, there are operators mixing the particle and antiparticle
states. Therefore the very notions of particle and
antiparticles are approximate, the conservation of electric,
baryon and lepton charges is also approximate and
superpositions of particle and antiparticle states are not
strictly prohibited. These notions and conservation laws are
valid only in the approximation when one considers only
transformations not mixing particle and antiparticle states. In
Section \ref{breaking} we discussed a possibility that one IR
can be split into independent IRs for a particle and its
antiparticle. It has been shown that such a possibility can be
implemented but only at the expense of breaking the exact AdS
symmetry. Since we accept that symmetry is the most important
criterion, we conclude that {\it the very notions of particle
and antiparticles are approximate and the electric, baryon and
lepton charges are only approximately conserved quantities}.
The non-conservation of the baryon and lepton quantum numbers
has been already considered in models of Grand Unification but
the electric charge has been always believed to be a strictly
conserved quantum number. The non-conservation of these quantum
numbers also completely changes the status of the problem known
as "baryon asymmetry of the Universe" since at early stages of
the Universe energies were much greater than now and therefore
transitions between particles and antiparticles had a much
greater probability.

We have also shown in Section \ref{S7} that in GFQT there can
be no neutral elementary particles. As explained in this
section, the spin-statistics theorem can be treated as a
requirement that standard quantum theory should be based on
complex numbers. This requirement excludes the existence of
neutral elementary particles. One might conclude that since in
GFQT the photon cannot be elementary, this theory cannot be
realistic and does not deserve attention. We believe however,
that the nonexistence of neutral elementary particles in GFQT
shows that the photon (and the graviton and the Higgs boson if
they exist) should be considered on a deeper level. In Section
\ref{Singletons} we argued that in GFQT a possibility that
massless particles are composite states of Dirac singletons is
even more attractive than in standard theory.

An important problem is what GFQT can say about supersymmetry.
There is no doubt that supersymmetry is a beautiful idea. On
the other hand, one might say that there is no reason for
nature to have both, elementary fermions and elementary bosons
since the latter can be constructed from the former. A well
know historical analogy is that the simplest covariant equation
is not the Klein-Gordon equation for spinless fields but the
Dirac and Weyl equations for the spin 1/2 fields since the
former is the equation of the second order while the latter are
the equations of the first order. In Section \ref{SS} we have
described results for modular IRs of the osp(1,4) superalgebra
and noted that supersymmetry does not impose strong
restrictions on the structure of modular IRs. Therefore the
problem of supersymmetry remains open.

Consider now the following very important question. If we
accept that the cosmological constant $\Lambda$ is positive
then in the framework of standard theory based on complex
numbers we have to draw a conclusion that the dS algebra
so(1,4) is a more relevant symmetry algebra than the Poincare
and AdS algebras. Therefore elementary particles should be described
by IRs of the so(1,4) algebra rather than IRs of the other two
algebras. As shown in Reference \cite{JPA}, the only
possible interpretation of IRs of the so(1,4) algebra is that
they describe a particle and its antiparticle simultaneously.
Therefore the very notions of particles and antiparticles are
only approximate and the electric, baryon and lepton charges
can be only approximately conserved quantities. In addition,
only fermions can be elementary (since in standard theory only
$\alpha{\bar\alpha}=1$ is possible while Equation (\ref{77}) is
not) and there are no neutral elementary particles. In view of
these remarks, a question arises whether the consideration of
modular IRs of the so(2,3) algebra is compatible with the fact
that $\Lambda>0$. In standard theory a difference between IRs
of the so(2,3) and so(1,4) algebras is that an IR of the
so(2,3) algebra where the operators $M^{\mu 5}$ ($\mu=0,1,2,3$)
are Hermitian can be treated as IRs of the so(1,4) algebra
where these operators are anti-Hermitian and vice versa. As
noted in Section \ref{S1}, in GFQT a probabilistic
interpretation is only approximate. Therefore one cannot
exclude a possibility that elementary particles can be
described by modular IRs discussed in this paper while modular
representations describing symmetry of macroscopic bodies at
cosmological distances are modular analogs of standard
representations of the so(1,4) algebra.

\section*{Acknowledgements}
The author is grateful to F. Coester, S. Dolgobrodov, H.
Doughty, C. Hayzelden, V. Netchitailo, M. Planat, W. Polyzou,
M. Saniga and T. Shtilkind for useful discussions.


\bibliographystyle{mdpi}
\makeatletter
\renewcommand\@biblabel[1]{#1. }
\makeatother

\end{document}